\documentclass[12pt]{article}
\pdfoutput=1

\usepackage[utf8]{inputenc}
\usepackage[left=2.55cm, right=2.55cm, top=2.55cm, bottom=2.55cm]{geometry}
\usepackage{amsmath,amssymb,amsbsy}
\usepackage{slashed}
\usepackage{xcolor}
\usepackage{graphicx}
\usepackage{url}
\usepackage{cancel}
\usepackage{cite}
\usepackage[colorlinks=true,allcolors=darkpurple,pdfborder={0 0 0},linktocpage=false,pdfencoding=auto]{hyperref}
\usepackage{tabularx,booktabs}
\usepackage{multicol}
\usepackage{units}
\usepackage{xspace}
\usepackage[labelfont=bf]{caption}
\usepackage[section]{placeins}
\usepackage{subcaption}
\usepackage{soul} 

\definecolor{darkred}{rgb}{0.6,0,0}
\definecolor{darkpurple}{rgb}{0.5,0,0.5}

\def\hc{\text{h.c.}}
\def\cc{\text{c.c.}}
\def\BR{\text{BR}}

\newcommand{\AddrIFIC}{%
Instituto de F\'{i}sica Corpuscular, CSIC-Universitat de Val\`{e}ncia, 46980 Paterna, Spain}

\newcommand{\AddrFISTEO}{%
Departament de F\'{\i}sica Te\`{o}rica, Universitat de Val\`{e}ncia, 46100 Burjassot, Spain}

\begin{document}

\vspace*{-2cm}
\begin{flushright}
IFIC/20-40 \\
\vspace*{2mm}
\end{flushright}

\begin{center}
\vspace*{15mm}

\vspace{1cm}
{\Large \bf 
Ultralight scalars in leptonic observables
} \\
\vspace{1cm}

{\bf Pablo Escribano$^{a}$, Avelino Vicente$^{a,b}$}

\vspace*{.5cm}
$(a)$ \AddrIFIC \\
$(b)$ \AddrFISTEO

 \vspace*{.3cm} 
\href{mailto:pablo.escribano@ific.uv.es}{pablo.escribano@ific.uv.es}, \href{mailto:avelino.vicente@ific.uv.es}{avelino.vicente@ific.uv.es}
\end{center}

\vspace*{10mm}

\begin{abstract}\noindent\normalsize
  Many new physics scenarios contain ultralight scalars, states which
  are either exactly massless or much lighter than any other massive
  particle in the model. Axions and majorons constitute well-motivated
  examples of this type of particle. In this work, we explore the
  phenomenology of these states in low-energy leptonic
  observables. After adopting a model independent approach that
  includes both scalar and pseudoscalar interactions, we briefly
  discuss the current limits on the diagonal couplings to charged
  leptons and consider processes in which the ultralight scalar $\phi$
  is directly produced, such as $\mu \to e \, \phi$, or acts as a
  mediator, as in $\tau \to \mu \mu \mu$. Contributions to the charged
  leptons magnetic and electric moments are studied
  as well.
\end{abstract}

\newpage

\tableofcontents

\newpage

\section{Introduction}
\label{sec:intro}

Lepton flavor physics is about to live a golden age. Several
state-of-the-art experiments recently started taking data and a few
more are about to begin~\cite{Calibbi:2017uvl}. These include new
searches for lepton flavor violating (LFV) processes, forbidden in the
Standard Model (SM), as well as more precise measurements of lepton
flavor conserving observables, such as charged lepton anomalous
magnetic moments. The search for LFV in processes involving charged
leptons is strongly motivated by the observation of LFV in the neutral
sector (in the form of neutrino flavor oscillations). In what concerns
muon observables, the search for the radiative LFV decay $\mu \to e
\gamma$ is going to be led by the second phase of the MEG experiment,
MEG-II~\cite{Baldini:2018nnn,Papa:2020umc}, while the long-awaited
Mu3e experiment will aim at an impressive sensitivity to branching
ratios for the 3-body decay $\mu \to eee$ as low as
$10^{-16}$~\cite{Berger:2014vba,Papa:2020umc}. A plethora of promising
experiments looking for neutrinoless $\mu-e$ conversion in nuclei is
also planned. Flavor factories and experiments aiming at a broad
spectrum of flavor observables, such as Belle II and LHCb, will also
contribute to this era of lepton flavor, mainly due to their high
sensitivities in the measurement of tau lepton
observables~\cite{Aushev:2010bq,Perez:2019cdy}. On the flavor conserving side,
improved measurements of the muon anomalous magnetic moment are
expected at the Muon g-2 experiment~\cite{Grange:2015fou}, hopefully
shedding light on a well-known long-standing experimental anomaly.

With such an exciting experimental perspective in the coming years, it
is natural to ask what type of new physics can be probed. In this work
we will concentrate on ultralight scalars that couple to charged
leptons and study their impact on leptonic observables. In this
context, we will use the term \textit{ultralight scalar} to refer to a
generic scalar $\phi$ that is much lighter than the electron, $m_\phi
\ll m_e$, and can therefore be produced on-shell in charged lepton
decays. In practice, this also means that $\phi$ can be assumed to be
approximately massless in all considered physical processes. We will
take a model independent approach and neglect $m_\phi$ in our
analytical calculations. Actually, this is not an approximation if
$\phi$ is exactly massless, the case for a Goldstone boson whose mass
is protected by a (spontaneously broken) global continuous symmetry.

There are many well-known examples of such ultralight scalars. If the
apparent absence of CP violation in the strong interactions is
explained by means of the Peccei-Quinn mechanism~\cite{Peccei:1977hh},
a new pseudoscalar state must exist: the
axion~\cite{Weinberg:1977ma,Wilczek:1977pj}. Although its mass is not
predicted and can vary over a wide range of
scales~\cite{DiLuzio:2020wdo}, a large fraction of the parameter space
(corresponding to large axion decay constants) leads to an ultralight
axion. Interestingly, such low mass axion would be of interest as a
possible component of the dark matter of the
Universe~\cite{Preskill:1982cy,Abbott:1982af,Dine:1982ah}. Axion-like
particles, or ALPs, generalize this type of scenario by making the
mass and decay constant two independent parameters. This allows for a
larger parameter space, again including a substantial portion with
very low ALP masses. The solution to the strong CP problem could also
be intimately related to the flavor problem of the
SM~\cite{Davidson:1981zd,Wilczek:1982rv}. This naturally leads to a flavored
axion~\cite{Calibbi:2016hwq,Ema:2016ops,Alanne:2018fns,CentellesChulia:2020bnf},
although an axion with flavor-blind interactions is also
possible~\cite{Reig:2018ocz}. Another popular ultralight scalar is the
majoron, the Goldstone boson associated to the breaking of global
lepton
number~\cite{Chikashige:1980ui,Gelmini:1980re,Schechter:1981cv,Aulakh:1982yn}. While
this state can gain a small mass by various mechanisms, and then be a
possible dark matter candidate~\cite{Heeck:2017kxw,Reig:2019sok}, it
is expected to be exactly massless in the absence of explicit breaking
of lepton number. Another possible ultralight scalar is the familon,
the Goldstone boson of spontaneously broken global family
symmetry. Finally, the Universe could also be filled with ultralight
scalars in the form of fuzzy cold dark matter~\cite{Hu:2000ke}.

While many of the previously discussed examples are pseudoscalar
states, the ultralight scalar $\phi$ can also have pure scalar
couplings. This would be the case for a massless Goldstone boson if
the associated global symmetry is non-chiral. Therefore, restricting
the phenomenological exploration to just pseudoscalars would miss a
relatively large number of well-motivated scenarios. This has actually
been the case in many recent
works~\cite{Bjorkeroth:2018dzu,Bjorkeroth:2018ipq,Gavela:2019wzg,Bauer:2019gfk,Bonnefoy:2019lsn,Cornella:2019uxs,Albrecht:2019zul,MartinCamalich:2020dfe,Endo:2020mev,Iguro:2020rby,Calibbi:2020jvd},
which were mainly interested in the phenomenology of flavored axions
(or ALPs) and majorons~\cite{Heeck:2019guh}.

Motivated by the principle of generality, we will consider a generic
scenario where the CP nature of $\phi$ is not determined and explore
several leptonic observables of interest. These include processes
in which $\phi$ is produced in the final state, such as $\ell_\alpha
\to \ell_\beta \, \phi$ or $\ell_\alpha \to \ell_\beta \, \phi \,
\gamma$. In this case, we will generalize previous results in the
literature, typically obtained for pure pseudoscalars or for the case
of a massive $\phi$. We will also study processes in which $\phi$ is
not produced, but acts as a mediator. A prime example of this category
is $\ell_\alpha^- \to \ell_\beta^- \ell_\beta^- \ell_\beta^+$. To the
best of our knowledge, the mediation of this process by an ultralight
axion has only been previously considered
in~\cite{Bjorkeroth:2018dzu}. We will extend the study to more general
scalar states and provide detailed analytical expressions for the
decay width of the process. The analogous $\ell_\alpha^- \to
\ell_\beta^- \ell_\gamma^- \ell_\gamma^+$ and $\ell_\alpha^- \to
\ell_\beta^+ \ell_\gamma^- \ell_\gamma^-$ decays will also be studied,
in this case for the first time here. Charged lepton anomalous
magnetic moments constitute other interesting examples of observables
induced by the ultralight $\phi$.

The rest of the manuscript is organized as follows. We introduce our
general setup, as well as our notation and conventions, in
Sec.~\ref{sec:lag}. In Sec.~\ref{sec:bounds} we discuss the current
bounds on the lepton flavor conserving couplings of the scalar
$\phi$. These are often constrained by studing their impact on
astrophysical processes, but also receive indirect bounds due to their
contribution to the 1-loop coupling of $\phi$ to photons, as we will
show. In Sec.~\ref{sec:obs} we discuss the impact of
  $\phi$ on several leptonic observables and derive analytical
  expressions for them. Phenomenological implications are considered
  in Sec.~\ref{sec:pheno}. We summarize our findings and conclude in
Sec.~\ref{sec:conclusions}. Finally, a pedagogical discussion on an
alternative parametrization of the $\phi$ Lagrangian in terms of
derivative interactions is provided in Appendix~\ref{sec:polar}.

\section{Effective Lagrangian}
\label{sec:lag}

We are interested in charged leptons processes taking place at low
energies in the presence of the ultralight real scalar $\phi$. For
practical purposes, we will consider $\phi$ to be exactly massless,
but our results are equally valid for a massive $\phi$, as long as
$m_\phi \ll m_e$ holds. The interaction of the scalar $\phi$ with a
pair of charged leptons $\ell_\alpha$ and $\ell_\beta$, with
$\alpha,\beta = e,\mu,\tau$, can be generally parametrized by
\begin{equation}
  \mathcal{L}_{\ell \ell \phi} = \phi \, \overline{\ell}_\beta \left( S_L^{\beta \alpha} P_L + S_R^{\beta \alpha} P_R \right) \ell_\alpha + \hc \, , \label{eq:lagS}
\end{equation}
where $P_{L,R} = \frac{1}{2} (1 \mp \gamma_5)$ are the usual chiral
projectors. No sum over the $\alpha$ and $\beta$ charged lepton flavor
indices is performed. $S_L$ and $S_R$ are dimensionless coefficients
and we consider all possible flavor combinations: $\beta \alpha =
\left\{ ee, \mu\mu,\tau\tau,e\mu,e\tau,
\mu\tau\right\}$. Eq.~\eqref{eq:lagS} describes the most general
effective interaction between the ultralight scalar $\phi$ and a pair
of charged leptons. In particular, we note that Eq.~\eqref{eq:lagS}
includes both scalar and pseudoscalar interactions as well as flavor
violating (charged lepton fields with $\alpha \ne \beta$) and flavor
conserving (charged lepton fields with $\alpha = \beta$)
interactions. An alternative parametrization for this Lagrangian based
on the introduction of derivative interactions, applicable to the case
of pseudoscalar interactions only, is discussed in
Appendix~\ref{sec:polar}.

Some of the LFV observables considered below receive contributions
from the usual dipole and 4-fermion operators. Therefore, our full
effective Lagrangian is given by
\begin{equation} \label{eq:lag}
  \mathcal{L} = \mathcal{L}_{\ell \ell \phi} + \mathcal{L}_{\ell \ell \gamma} + \mathcal{L}_{4 \ell} \, ,
\end{equation}
with
\begin{align}
  \mathcal{L}_{\ell \ell \gamma} & = \frac{e \, m_\alpha}{2} \, \overline{\ell}_{\beta} \, \sigma^{\mu \nu} \left[ \left( K_2^L \right)^{\beta \alpha} P_L + \left( K_2^R \right)^{\beta \alpha} P_R \right] \ell_{\alpha} F_{\mu \nu} + \hc \, , \label{eq:lagA} \\
  \mathcal{L}_{4 \ell} & = \sum_{I=S, V, T \atop X, Y=L, R} \left( A_{X Y}^{I} \right)^{\beta \alpha \delta \gamma} \, \overline{\ell}_\beta \Gamma_I P_X \ell_{\alpha} \, \overline{\ell}_\delta \Gamma_I P_Y \ell_\gamma + \hc \, , \label{eq:lag4F}  
\end{align}
where $F_{\mu \nu} = \partial_\mu A_\nu - \partial_\nu A_\mu$ is the
electromagnetic field strength tensor, with $A_\mu$ the photon field,
and we have defined $\Gamma_S = 1$, $\Gamma_V = \gamma_{\mu}$ and
$\Gamma_T = \sigma_{\mu \nu}$. No sum over the $\alpha$, $\beta$,
$\gamma$ and $\delta$ charged lepton flavor indices is performed in
Eqs.~\eqref{eq:lagA} and \eqref{eq:lag4F}. The coefficients $K_2^X$
and $A_{X Y}^{I}$, with $I = S,V,T$ and $X,Y = L,R$, have dimensions
of mass$^{-2}$.  We assume $m_\alpha > m_\beta$ and therefore
normalize the Lagrangian in Eq.~\eqref{eq:lagA} by including the mass
of the heaviest charged lepton in the process of
interest. Eq.~\eqref{eq:lagA} contains the usual photonic dipole
operators, which contribute to $\ell_\alpha \to \ell_\beta \gamma$ and
lead to
\begin{equation} \label{eq:widthLLG}
  \Gamma \left( \ell_\alpha \to \ell_\beta \gamma \right) = \frac{e^2 \, m_\alpha^5}{16 \, \pi} \left[ \left|\left(K_2^L\right)^{\beta\alpha}\right|^2 + \left|\left(K_2^R\right)^{\beta \alpha}\right|^2 \right] \, ,
\end{equation}
while Eq.~\eqref{eq:lag4F} contains 4-lepton operators.  In summary,
the effective Lagrangian in Eq.~\eqref{eq:lag} corresponds to the one
in~\cite{Porod:2014xia}, extended to include the new operators with
the scalar $\phi$ introduced in Eq.~\eqref{eq:lagS}.

In the following, we will disregard $\phi$ interactions with quarks
and concentrate on purely leptonic observables, such as the LFV decays
$\ell_\alpha \to \ell_\beta \, \phi$ or $\ell_\alpha \to \ell_\beta
\ell_\beta \ell_\beta$, and the electron and muon anomalous magnetic
and electric dipole moments. Even though $\phi$
couplings to quarks are possible, and indeed present in specific
realizations of our general scenario, the prime example being the QCD
axion, they introduce a large model dependence.  We also note that
leptophilic ultralight scalars, such as the majoron, are also
well-motivated possibilities that naturally appear in models with
spontaneous violation of global lepton number.

\section{Bounds on lepton flavor conserving couplings}
\label{sec:bounds}

Let us comment on the current experimental contraints on the lepton
flavor conserving couplings of the scalar $\phi$. We will start
discussing the stellar cooling mechanism. Since this subject has been
extensively studied in the literature, and we do not want to delve
further into the topic, only a brief outline will be presented. Then
we will discuss another source of constraints, the 1-loop coupling
between $\phi$ and a pair of photons.

\subsection{Stellar cooling}
\label{subsec:stellar}

The production of $\phi$ scalar particles inside stars, followed by
their emission, may constitute a powerful stellar cooling
mechanism. If this process takes place at a high rate, it may alter
star evolution, eventually leading to conflict with astrophysical
observations~\cite{Raffelt:1994ry}. This allows one to place strong
constraints on the $\phi$ scalar couplings. The dominant cooling
mechanisms are scalar bremsstrahlung in lepton-nucleus scattering,
$\ell^- + N \rightarrow \ell^- + N + \phi$, and the Compton process
$\gamma + \ell^- \rightarrow \ell^- + \phi$. Their relative importance
depends on the density and temperature of the medium, and therefore on
the astrophysical scenario. In particular, the Compton process
dominates only at low densities and high temperatures, conditions that
can be found in red giants. Limits can also be derived from the
production of ultralight scalars in supernovae. The scalar $\phi$ can
be efficiently produced and, since it will typically escape without
interacting with the medium, a net transport of energy out of the
supernova will take place. Such a loss of energy may dramatically
affect other processes taking place in the supernova, such as neutrino
production.

Plenty of works have recently studied the question of cooling by the
emission of ultralight scalars in astrophysical
scenarios~\cite{DiLuzio:2020wdo,Bollig:2020xdr,Calibbi:2020jvd,Croon:2020lrf,DiLuzio:2020jjp}.
However, to the best of our knowledge, all of them consider axions or
ALPs. These are low-mass pseudoscalars and thus, their impact on
stellar evolution can only be used to constrain pseudoscalar
couplings. Even though we will not provide a detailed calculation to
support this statement, we will argue that similar bounds can be set
on the scalar couplings.

To make explicit the pure scalar and pseudoscalar interactions, we can
use a redefinition of our Lagrangian in Eq.~\eqref{eq:lagS} which, for
the diagonal terms, can be written as
\begin{equation}
  \mathcal{L}_{\ell \ell \phi}^{\rm diag} = \phi \, \overline{\ell}_\beta \left( S^{\beta \beta} P_L + S^{\beta \beta \ast} P_R \right) \ell_\beta = \phi \, \overline{\ell}_\beta \left[ \text{Re} \, S^{\beta \beta} - i \, \text{Im} \, S^{\beta \beta} \, \gamma_5 \right] \ell_\beta \, , \label{eq:diag}
\end{equation}
with $S^{\beta \beta} = S_L^{\beta \beta} + S_R^{\beta \beta \ast}$. For a pure pseudoscalar, only $\text{Im} \,
S^{\beta \beta}$ is present.

The currently most stringent limit on the pseudoscalar coupling with
electrons is obtained from white dwarfs. Specifically, the limit is
obtained by considering the bremsstrahlung process, which can be very
efficient in the dense core of a white dwarf. Using data from the
Sloan Digital Sky Survey and the SuperCOSMOS Sky Survey,
Ref.~\cite{Calibbi:2020jvd} found (at 90\% C.L.)
\begin{equation} \label{eq:See}
  \text{Im} \, S^{e e} < 2.1 \times 10^{-13} \, .
\end{equation}
The coupling with muons has been recently studied in some
works~\cite{Bollig:2020xdr,Calibbi:2020jvd,Croon:2020lrf}. In this
case the process ultimately used to set the contraint is neutrino
production, clearly suppressed if energy is transported out of the
supernova by scalars produced in $\mu + \gamma \to \mu + \phi$. Using
the famous supernova SN1987A, Ref.~\cite{Croon:2020lrf} has found
\begin{equation} \label{eq:Smumu}
  \text{Im} \, S^{\mu \mu} < 2.1 \times 10^{-10} \, .
\end{equation}

Setting precise limits for the scalar parts of the couplings would
imply the calculation of the cross sections and the energy-loss rates
per unit mass, as required to perform a complete analysis. Instead,
one can gauge the relevance of the bounds on the scalar couplings with
the following arguments. First, we note that if the charged lepton
mass is neglected, the scalar and pseudoscalar couplings contribute in
exactly the same way to the relevant cross sections. This is, however,
a bad approximation, due to the low energies involved in the
astrophysical scenarios that set the limits. For this reason, one must
keep the charged lepton mass. We have numerically integrated the cross
sections for a wide range of low energies and found that, for the same
numerical value of $\text{Re} \, S$ and $\text{Im} \, S$, the
scalar interaction always gives larger cross sections. Therefore, the
constraints on the scalar couplings will be stronger and we can
conclude that
\begin{equation}
  \text{Re} \, S^{\beta \beta} \lesssim \left[ \text{Im} \, S^{\beta \beta} \right]_{\max} \, ,
\end{equation}
with $\beta = e,\mu$. Nevertheless, we point out that a detailed
analysis of the cooling mechanism with pure scalars is required to
fully determine the corresponding bounds.

Finally, one should note that these limits are based on the
(reasonable) assumption that the scalar properties are not altered in
the astrophysical medium. In particular, its mass and couplings are
assumed to be the same as in vacuum. Some mechanisms have been
recently proposed~\cite{Bloch:2020uzh,DeRocco:2020xdt} (see also
previous work in~\cite{Masso:2006gc}) that would make this assumption
invalid. These works are mainly motivated by the recent XENON1T
results, which include a $3.5 \, \sigma$ excess of low-energy electron
recoil events~\cite{Aprile:2020tmw}. An axion explaining this excess
would violate the astrophysical constraints, since the required
coupling to electrons would be larger than the limit in
Eq.~\eqref{eq:See}, see for instance~\cite{DiLuzio:2020jjp}. This
motivates the consideration of mechanisms that alter the effective
couplings to electrons or the axion mass in high density scenarios. If
any of these mechanisms are at work, larger diagonal couplings would
be allowed. However, we note that additional bounds, not derived from
astrophysical observations, can be set on the diagonal couplings. This
is precisely what we proceed to discuss.

\subsection{1-loop coupling to photons}
\label{subsec:agg}

\begin{figure}[tb]
  \centering
  \includegraphics[width=0.5\linewidth]{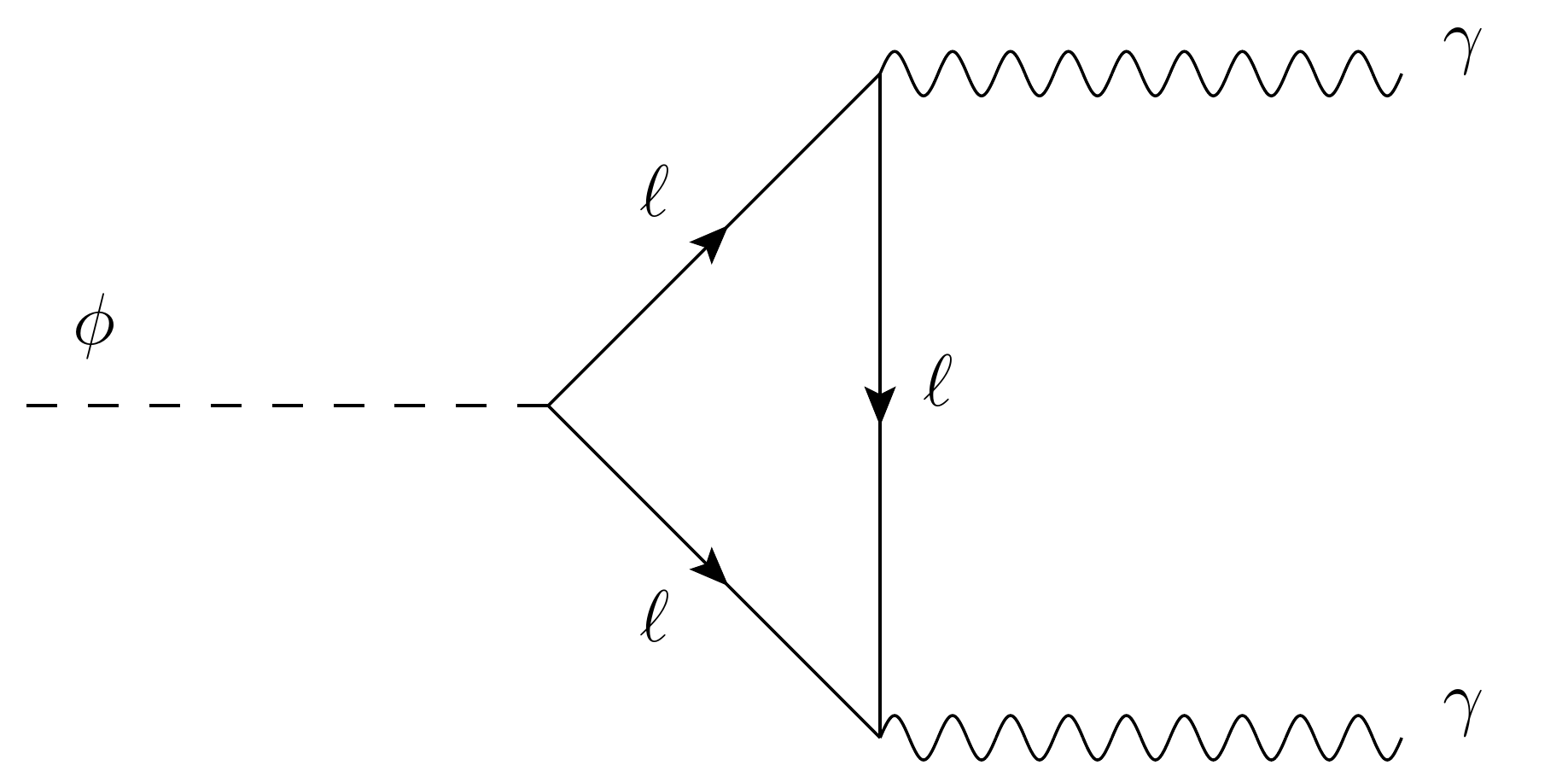}
  \caption{Loop induced coupling of $\phi$ to a pair of photons.}
  \label{fig:one-loop_phi_gamma_gamma}
\end{figure}

The interaction of the scalar $\phi$ to a pair of photons is described
by the effective Lagrangian
\begin{equation} \label{eq:LagSgammagamma}
  \mathcal{L}_{\phi \gamma \gamma} = g_{S \gamma \gamma} \, \phi \, F_{\mu \nu} F^{\mu \nu} + g_{A \gamma \gamma} \, \phi \, F_{\mu \nu} \widetilde{F}^{\mu \nu} \, ,
\end{equation}
where $g_{S \gamma \gamma}$ and $g_{A \gamma \gamma}$ are the
couplings for a pure scalar and a pure pseudoscalar, respectively, and
$\widetilde{F}^{\mu \nu}$ is the dual electromagnetic tensor, defined
as
\begin{equation}
  \widetilde{F}^{\mu \nu} = \frac{1}{2} \, \varepsilon^{\mu \nu \alpha \beta} \, F_{\alpha \beta} \, .
\end{equation}

The $g_{S \gamma \gamma}$ and $g_{A \gamma \gamma}$ couplings can be
induced at the 1-loop level from diagrams involving charged leptons,
as shown in Fig.~\ref{fig:one-loop_phi_gamma_gamma}. Since $g_{S
  \gamma \gamma}$ and $g_{A \gamma \gamma}$ are constrained by a
variety of experimental sources, this can be used to set indirect
constraints on the $\phi$ couplings to charged leptons introduced in
Eq.~\eqref{eq:lagS}. In particular, we will take advantage of this
relation to get additional limits on the lepton flavor conserving
couplings of $\phi$.

The 1-loop analytical expression for $g_{S \gamma \gamma}$ and $g_{A
  \gamma \gamma}$ can be written as~\cite{Djouadi:2005gj}
\begin{equation}
	\left| g_{I \gamma \gamma} \right|^2 = \frac{\alpha^2}{64 \, \pi^2} \, \left| \sum_\beta \, \frac{g_{I \beta \beta}}{m_\beta} \, A_{1 / 2}^I \left( \tau_\beta \right) \right|^2,
\end{equation}
where $I = S, A$ and we sum over $\beta = e, \mu,
  \tau$. Here $g_{I \beta \beta}$ denote the $\phi$ couplings to the
  charged leptons, and their relation to $S_L$ and $S_R$ is given
  below. $A_{1 / 2}^{S}$ and $A_{1 / 2}^{A}$ are 1-loop fermionic
functions defined as
\begin{equation}
  A_{1 / 2}^{S}(\tau_\beta) = 2 \left[ \tau_\beta + \left( \tau_\beta - 1 \right) f \left( \tau_\beta \right) \right] \tau_\beta^{-2}
\end{equation}
for the scalar coupling and
\begin{equation}
  A_{1 / 2}^{A}(\tau_\beta) = 2 \tau_\beta^{-1} f \left( \tau_\beta \right)
\end{equation}
for the pseudoscalar case, with $\tau_\beta = m_\phi^2 / 4
m_\beta^2$. The function $f \left( \tau \right)$ can be found for
instance in~\cite{Carmi:2012in}. It is given by
\begin{equation}
  f(\tau) \equiv\left\{\begin{array}{ll}
  \arcsin ^{2} \sqrt{\tau} & \tau \leq 1 \\
  -\frac{1}{4}\left[\log \frac{1+\sqrt{1-\tau^{-1}}}{1-\sqrt{1-\tau^{-1}}}-i \pi\right]^{2} & \tau>1
  \end{array}\right. .
\end{equation}
In this work we consider the case of an ultralight scalar. In the
massless limit, the loop functions reduce simply to $A_{1/2}^S
\left(0\right) = \frac{4}{3}$ and $A_{1/2}^A \left(0\right) = 2$, and
then we can write
\begin{equation}
  \begin{split}
    \left| g_{S \gamma \gamma} \right|^2 = \frac{\alpha^2}{36 \, \pi^2} \, \left| \sum_\beta \, \frac{g_{S \beta \beta}}{m_\beta} \right|^2, \\
    \left| g_{A \gamma \gamma} \right|^2 = \frac{\alpha^2}{16 \, \pi^2} \, \left| \sum_\beta \, \frac{g_{A \beta \beta}}{m_\beta} \right|^2,
  \end{split}
\end{equation}
with the couplings to the charged leptons being given by
\begin{equation}
  \begin{split}
    g_{S \beta \beta} = \text{Re} \, S^{\beta \beta} \, , \\
    g_{A \beta \beta} = \text{Im} \, S^{\beta \beta} \, .
  \end{split}
\end{equation}

We are now in position to compare to the current experimental limits
on the coupling to photons. These are of two types. First, let us
consider astrophysical limits. Magnetic fields around astrophysical
sources of photons may transform these into scalars, an effect that
can be used to set constraints on their
coupling. Ref.~\cite{Antoniou:2015sga} provides a comprehensive
recollection of limits from astrophysical observations. Using results
from~\cite{Burrage:2009mj}, this reference finds that for scalar
masses in the range $m_\phi \ll 1 \, \text{peV} \, - \, 1 \,
\text{neV}$, astrophysical constraints imply
\begin{equation}
  g_{I \gamma \gamma} \lesssim \left( 10^{-12} - 10^{-11} \right) \, \text{GeV}^{-1}
\end{equation}
for both scalar and pseudoscalar couplings. Taking this into account,
we can find the relations
\begin{equation}
  \begin{split}
    \left| \sum_\beta \, \frac{\text{Re} \, S^{\beta \beta}}{m_\beta} \right|^2 = \frac{36 \, \pi^2}{\alpha^2} \, g_{S \gamma \gamma}^2 < 6.7 \times 10^{-16} \, \text{GeV}^{-2} \, , \\
    \left| \sum_\beta \, \frac{\text{Im} \, S^{\beta \beta}}{m_\beta} \right|^2 = \frac{16 \, \pi^2}{\alpha^2} \, g_{A \gamma \gamma}^2 < 3.0 \times 10^{-16} \, \text{GeV}^{-2} \, ,
  \end{split}
\end{equation}
which translate into very stringent bounds on the diagonal couplings
to charged leptons, $S^{ee} \lesssim 10^{-11}$ and $S^{\mu\mu} \lesssim
10^{-9}$. The OSQAR experiment~\cite{Ballou:2014myz}, a
light-shining-through-a-wall experiment, has also derived limits for
massless scalars. Again, these are valid for both scalar and
pseudoscalar couplings,
\begin{equation}
  g_{I \gamma \gamma} < 5.76 \times 10^{-8} \, \text{GeV}^{-1} \, ,
\end{equation}
and therefore,
\begin{equation}
  \begin{split}
    \left| \sum_\beta \, \frac{\text{Re} \, S^{\beta \beta}}{m_\beta} \right|^2 = \frac{36 \, \pi^2}{\alpha^2} \, g_{S \gamma \gamma}^2 < 3.8 \times 10^{-8} \, \text{GeV}^{-2} \, , \\
    \left| \sum_\beta \, \frac{\text{Im} \, S^{\beta \beta}}{m_\beta} \right|^2 = \frac{16 \, \pi^2}{\alpha^2} \, g_{A \gamma \gamma}^2 < 1.7 \times 10^{-8} \, \text{GeV}^{-2} \, .
  \end{split}
\end{equation}
These relations also imply strong contraints on the diagonal couplings
to charged leptons, but milder than in the previous case, $S^{ee}
\lesssim 10^{-7}$ and $S^{\mu\mu} \lesssim 10^{-5}$.

Finally, we point out that these indirect limits are strictly only
valid if the diagrams in Fig.~\ref{fig:one-loop_phi_gamma_gamma} are
the only contribution to the $\phi$ coupling to photons. If more
contributions exist, possible cancellations among them may reduce the
total coupling so that the constraints are satisfied for larger
couplings to charged leptons. We should also note that astrophysical
constraints are subject to the same limitation discussed above. They
rely on the assumption that the properties of $\phi$ in the
astrophysical medium are the same as in vacuum.

\section{Leptonic observables}
\label{sec:obs}
               
\subsection[\texorpdfstring{$\ell_\alpha \to \ell_\beta \, \phi$}{\unichar{"2113}\unichar{"03B1} \unichar{"2192} \unichar{"2113}\unichar{"03B2} \unichar{"03D5}}]{$\boldsymbol{\ell_\alpha \to \ell_\beta \, \phi}$}
\label{sec:obsbetaphi}

The off-diagonal $S_A^{\beta \alpha}$ scalar couplings, with $A =
L,R$, can be directly constrained by the LFV decays $\ell_\alpha \to
\ell_\beta \, \phi$. Using the effective Lagrangian in
Eq.~\eqref{eq:lagS}, it is straightforward to obtain
\begin{equation}
  \Gamma \left(\ell_\alpha \to \ell_\beta \, \phi \right) = \frac{m_\alpha}{32 \, \pi} \left( \left| S_L^{\beta \alpha} \right|^2 + \left| S_R^{\beta \alpha} \right|^2 \right) \, ,
  \label{eq:decaywidth_betaphi}
\end{equation}
where terms proportional to the small ratio $m_\beta / m_\alpha$ have
been neglected.~\footnote{We must notice that this approximation is
  not equally good for all $\ell_\alpha \to \ell_\beta \, \phi$
  cases. This is because the ratio $m_\mu / m_\tau \sim 0.1$ is not
  completely negligible. Therefore, while the approximation is very
  good for $\mu \to e \, \phi$ and $\tau \to e \, \phi$, it may lead
  to an error of the order of $20 \%$ in $\tau \to \mu \, \phi$. This
  deviation is acceptable, but can be accounted for by including
  additional terms proportional to $m_\mu / m_\tau$, hence leading to
  a much more complicated analytical expression. Completely analogous
  comments can be made for the rest of the observables discussed in
  this Section.}

\subsection[\texorpdfstring{$\ell_\alpha \to \ell_\beta \, \gamma \, \phi$}{\unichar{"2113}\unichar{"03B1} \unichar{"2192} \unichar{"2113}\unichar{"03B2} \unichar{"03B3} \unichar{"03D5}}]{$\boldsymbol{\ell_\alpha \to \ell_\beta \, \gamma \, \phi}$}

The decay width for the 3-body LFV process $\ell_\alpha \to \ell_\beta
\, \gamma \, \phi$ can be written as
\begin{equation}
  \Gamma \left( \ell_\alpha \to \ell_\beta \, \gamma \, \phi \right) = \frac{\alpha \, m_\alpha}{64 \pi^2} \left( \left| S^{\beta \alpha}_L \right|^2 + \left| S^{\beta \alpha}_R \right|^2 \right) \mathcal{I} \left( x_{ \min } , y_{ \min } \right) \,,
  \label{eq:decaywidth_betagammaphi}
\end{equation}
where terms proportional to $m_\beta / m_\alpha$ have been
neglected. Here $\mathcal{I} \left( x_{ \min } , y_{ \min } \right)$
is a phase space integral given by
\begin{equation}
  \mathcal{I} \left( x_{ \min } , y_{ \min } \right) = \int \text{d} x \, \text{d} y \, \frac{ \left( x - 1 \right) \left( 2 - x y - y \right) }{ y^{2} \left( 1 - x - y \right) } \, ,
  \label{eq:phase_space_integral}
\end{equation}
and we have introduced the usual dimensionless parameters $x$ and
$y$, defined as
\begin{equation}
  x = \frac{2 E_{\beta}}{m_{\alpha}} \quad, \quad y = \frac{2 E_{\gamma}}{m_{\alpha}} \, ,
\end{equation}
which, together with $z = 2 E_{\phi} / m_{\alpha}$, must fulfill the
kinematical condition $x + y + z = 2$. We point out that our
analytical results match those in~\cite{Hirsch:2009ee}, except for
redefinitions in the couplings.~\footnote{In the model considered
  in~\cite{Hirsch:2009ee}, the right-handed coupling was suppressed
  and hence neglected.}

The phase space integral in Eq.~\eqref{eq:phase_space_integral}
depends on $x_{\min}$ and $y_{\min}$, the minimal values that the $x$
and $y$ parameters may take. While one could naively think that these
are just dictated by kinematics, they are actually determined by the
minimal $\ell_\beta$ lepton and photon energies measured in a given
experiment. This not only properly adapts the calculation of the phase
space integral to the physical region explored in a real experiment,
but also cures the kinematical divergences that would otherwise
appear. In fact, we note that the integral in
Eq.~\eqref{eq:phase_space_integral} diverges when the photon energy
vanishes ($y \to 0$). This is the well-known infrared divergence that
also appears, for instance, in the radiative SM decay $\mu \to e \nu
\bar \nu \gamma$. Another divergence is encountered when the photon
and the $\ell_\beta$ lepton in the final state are emitted in the same
direction. The angle between their momenta is given by
\begin{equation} \label{eq:theta}
\cos \theta_{\beta \gamma} = 1 + \frac{2 - 2(x+y)}{xy} \, .
\end{equation}
Since we work in the limit $m_\beta = 0$, one finds a colinear
divergence in configurations in which the photon and the $\ell_\beta$
lepton have their momenta aligned ($\theta_{\beta \gamma} \to
0$). However, any real experimental setup has a finite experimental
resolution, which implies a non-zero minimum measurable $E_\gamma$ and
a non-zero minimum $\theta_{\beta \gamma}$ angle. Therefore, by
restricting the phase space integration to the kinematical region
explored in a practical situation, all divergences disappear.

Direct comparison with
Eq.~\eqref{eq:decaywidth_betaphi} allows one to establish the relation
\begin{equation}
  \Gamma \left( \ell_\alpha \to \ell_\beta \, \gamma \, \phi \right) = \frac{\alpha}{2 \pi} \, \mathcal{I} \left( x_{ \min } , y_{ \min } \right) \Gamma \left( \ell_\alpha \to \ell_\beta \, \phi \right) \, ,
\end{equation}
which tells us that $\ell_\alpha \to \ell_\beta \, \gamma \, \phi$ is
suppressed with respect to $\ell_\alpha \to \ell_\beta \, \phi$ due to
an additional $\alpha$ coupling and a phase space factor. In fact, the
latter turns out to be the main source of suppression.

\subsection[\texorpdfstring{$\ell_\alpha \to \ell_\beta \, \gamma$}{\unichar{"2113}\unichar{"03B1} \unichar{"2192} \unichar{"2113}\unichar{"03B2} \unichar{"03B3}}]{$\boldsymbol{\ell_\alpha \to \ell_\beta \, \gamma}$} \label{sec:obsbetagamma}

\begin{figure}[!t]
  \centering
  \includegraphics[width=0.6\linewidth]{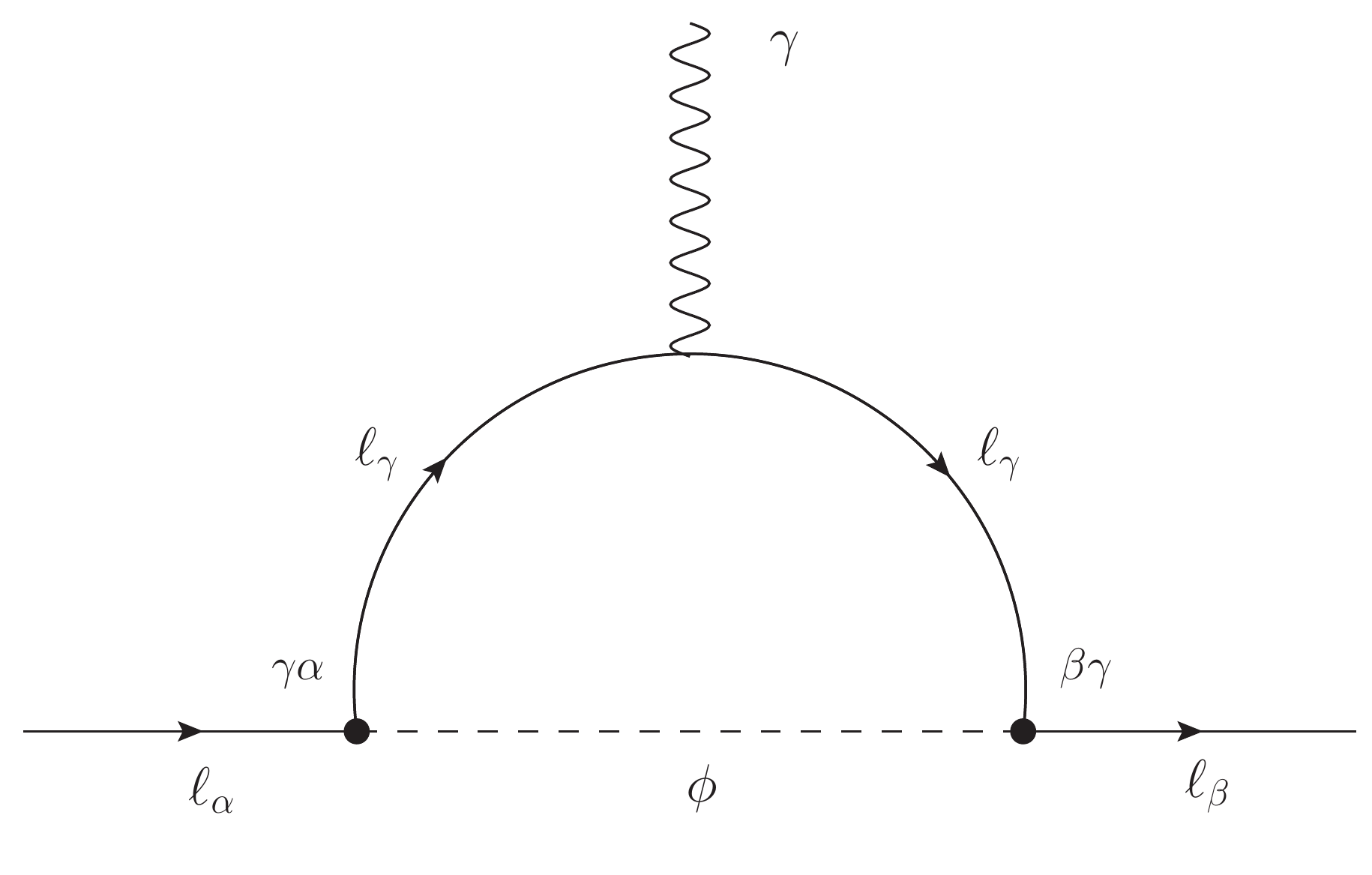}
  \caption{One-loop Feynman diagram contributing to the process $\ell_{\alpha} \to \ell_{\beta} \gamma$ described by the effective Lagrangian in Eq.~\eqref{eq:lagS}. The flavor indices of the couplings contributing to the diagram have been specified in the vertices.
  \label{fig:Diagrambeta_gamma}}
\end{figure}

The amplitude for the $\ell_\alpha \to \ell_\beta \,
  \gamma$ radiative decay only receives contributions from dipole
  operators and takes the general form
\begin{equation}
  \mathcal{M}_\phi = - e \, \bar{u}_{\beta}\left\{ m_{\alpha} \, \sigma^{\mu \nu} q_{\nu}\left[ \left(K_{2}^{L}\right)^{\beta \alpha} P_{L}+\left(K_{2}^{R}\right)^{\beta \alpha} P_{R}\right]\right\} u_\alpha \, \varepsilon^*_\mu \, ,
\end{equation}
where $u$ and $v$ are spinors and $q_\mu$ and $\varepsilon_\mu$ are the photon
  4-momentum and polarization vector, respectively. The $K_2^L$ and
  $K_2^R$ coefficients are induced at the one-loop level, as shown in
  Fig.~\ref{fig:Diagrambeta_gamma}. We find the expressions
\begin{equation} \label{eq:K2L}
    \left(K_2^L\right)^{\beta \alpha} = \frac{C_L^{\gamma \alpha} \, C_L^{\beta \gamma} \, f_1 \left( m_\alpha, m_\beta, m_\gamma \right) + C_L^{\gamma \alpha} \, C_R^{\beta \gamma} \, f_2 \left( m_\alpha, m_\beta, m_\gamma \right) + C_R^{\gamma \alpha} \, C_L^{\beta \gamma} \, f_2 \left( m_\beta, m_\alpha, m_\gamma \right)}{32 \pi^2 \, m_\alpha^5 \, m_\beta^3 \left(m_\alpha + m_\beta \right) \left( m_\alpha - m_\beta \right)^2 }, 
\end{equation}
\begin{equation} \label{eq:K2R}
    \left(K_2^R\right)^{\beta \alpha} = \frac{C_R^{\gamma \alpha} \, C_R^{\beta \gamma} \, f_1 \left( m_\beta, m_\alpha, m_\gamma \right) + C_L^{\gamma \alpha} \, C_R^{\beta \gamma} \, f_2 \left( m_\beta, m_\alpha, m_\gamma \right) + C_R^{\gamma \alpha} \, C_L^{\beta \gamma} \, f_2 \left( m_\alpha, m_\beta, m_\gamma \right)}{32 \pi^2 \, m_\alpha^5 \, m_\beta^3 \left(m_\alpha + m_\beta \right) \left( m_\alpha - m_\beta \right)^2 },
\end{equation}
where a sum over $\gamma$ is implicit here and the $f_i$ loop functions are defined as
\begin{equation}
  \begin{split}
    & f_1 \left( m_\alpha, m_\beta, m_\gamma \right) = 2 \, m_\alpha \, m_\beta \, m_\gamma \left[ m_\beta^2 \left(m_\alpha^2 - m_\beta^2 \right) \left(m_\alpha^2 - m_\gamma^2 \right) \log \frac{m_\gamma^2}{m_\gamma^2 - m_\alpha^2}  \right. \\
    & \left. + \, m_\alpha^2 \left(m_\beta^2 - m_\alpha^2 \right) \left(m_\beta^2 - m_\gamma^2 \right) \log \frac{m_\gamma^2}{m_\gamma^2 - m_\beta^2} + m_\alpha^2 \, m_\beta^2 \left( m_\alpha^2 - m_\beta^2 \right)^2 C_0 \left(0, m_\alpha^2, m_\beta^2, m_\gamma, m_\gamma, 0 \right) \right] \, ,
  \end{split}
\end{equation}
\begin{equation}
  \begin{split}
    f_2 \left( m_\alpha, m_\beta, m_\gamma \right) = & - m_\alpha^3 \, m_\beta^2 \left( m_\alpha^2 - m_\beta^2 \right) \left(m_\beta^2 + m_\gamma^2 \right) - m_\alpha \, m_\beta^4 \left( m_\alpha^4 - m_\gamma^4 \right)  \log \frac{m_\gamma^2}{m_\gamma^2 - m_\alpha^2}  \\
    & + \, m_\alpha^3 \left( m_\beta^2 - m_\gamma^2 \right) \left[ 2 \, m_\beta^2 \, m_\gamma^2 + m_\alpha^2 \left( m_\beta^2 - m_\gamma^2 \right) \right] \log \frac{m_\gamma^2}{m_\gamma^2 - m_\beta^2} \\
    & - \, 2 \, m_\alpha^3 \, m_\beta^4 \, m_\gamma^2 \left(m_\alpha^2 - m_\beta^2 \right) C_0 \left(0, m_\alpha^2, m_\beta^2, m_\gamma, m_\gamma, 0 \right) \, ,
  \end{split}
\end{equation}
and we have introduced here the usual scalar Passarino-Veltman
  three-point function
\begin{equation}
  C_0 \left(0, m_\alpha^2, m_\beta^2, m_\gamma, m_\gamma, 0 \right) = \frac{1}{2 \left(m_\alpha^2 - m_\beta^2 \right)} \Biggl[ \log^2 \left(- \frac{m_\gamma^2}{m_\alpha^2} \right) - \log^2 \left(- \frac{m_\gamma^2}{m_\beta^2} \right) + 2 \, \text{Li}_2 \, \frac{m_\gamma^2}{m_\alpha^2} - 2 \, \text{Li}_2 \, \frac{m_\gamma^2}{m_\beta^2} \Biggr] \, .
\end{equation}
The $C_{L,R}$ couplings that appear in
  Eqs.~\eqref{eq:K2L} and \eqref{eq:K2R} are related to the
  $S_{L,R}$ couplings introduced in the effective Lagrangian
  in Eq.~\eqref{eq:lagS}. The relation depends on the particular
  diagram under consideration:
\begin{equation}
  C_{L}^{\eta \rho} = \left\{ \begin{matrix}
                                     S_L^{\eta \rho} & m_\eta < m_\rho \\
                                     S_R^{\rho \eta *} & m_\eta > m_\rho \\
                                     S^{\eta \eta} & \eta = \rho                    
                                   \end{matrix} \right. \, ,
\end{equation}
and
\begin{equation}
  C_{R}^{\eta \rho} = \left\{ \begin{matrix}
                                     S_R^{\eta \rho} & m_\eta < m_\rho \\
                                     S_L^{\rho \eta *} & m_\eta > m_\rho \\
                                     S^{\eta \eta *} & \eta = \rho                    
                                   \end{matrix} \right. \, .
\end{equation}
It proves convenient to find approximate expressions
  for the $K_2^{L,R}$ coefficients, obtained at leading order in
  $m_\beta$. We find
\begin{equation}
  \left(K_2^L\right)^{\beta \alpha} = \frac{S^{\beta \beta} \, S_R^{\beta \alpha}}{32 \pi^2 \, m_\alpha^2} - \frac{S_L^{\beta \alpha} \left[ S^{\alpha \alpha} \left( \pi^2 - 6 \right) + S^{\alpha \alpha *} \left( \pi^2 - 9 \right) \right] }{96 \pi^2 \, m_\alpha^2} + \frac{1}{32 \pi^2 \, m_\alpha} \left\{ \begin{matrix}
                                     \frac{S_L^{\beta \gamma} \, S_R^{\gamma \alpha}}{m_\alpha} & m_\alpha \gg m_\gamma \gg m_\beta \\
                                     \frac{S_R^{\gamma \beta *} \, S_R^{\gamma \alpha}}{m_\alpha} & m_\alpha \gg m_\beta \gg m_\gamma \\
                                     - \frac{S_L^{\beta \gamma} \, S_R^{\alpha \gamma *}}{m_\gamma} & m_\gamma \gg m_\alpha                                    
                                   \end{matrix} \right.
\end{equation}
and
\begin{equation}
  \left(K_2^R\right)^{\beta \alpha}  = \frac{S^{\beta \beta *} \, S_L^{\beta \alpha}}{32 \pi^2 \, m_\alpha^2} - \frac{S_R^{\beta \alpha} \left[ S^{\alpha \alpha} \left( \pi^2 - 9 \right) + S^{\alpha \alpha *} \left( \pi^2 - 6 \right) \right]}{96 \pi^2 \, m_\alpha^2} + \frac{1}{32 \pi^2 \, m_\alpha} \left\{ \begin{matrix}
                                     \frac{S_R^{\beta \gamma} \, S_L^{\gamma \alpha}}{m_\alpha} & m_\alpha \gg m_\gamma \gg m_\beta \\
                                     \frac{S_L^{\gamma \beta *} \, S_L^{\gamma \alpha}}{m_\alpha} & m_\alpha \gg m_\beta \gg m_\gamma \\
                                     - \frac{S_R^{\beta \gamma} \, S_L^{\alpha \gamma *}}{m_\gamma} & m_\gamma \gg m_\alpha                                    
                                   \end{matrix} \right. ,
\end{equation}
We note, however, that these approximate expressions may only serve as an estimate for the order of magnitude of the $K_2^{L,R}$ coefficients, since large errors ($\sim 50 \%$) are obtained in some cases due to the appearance of large logs. Finally, upon substitution in Eq.~\eqref{eq:widthLLG}, one obtains the total decay width of the process.~\footnote{For completeness, we note that the expression for the $\ell_\alpha \to \ell_\beta \, \gamma$ decay width without neglecting $m_\beta^2$ is
\begin{equation*}
  \Gamma \left( \ell_\alpha \to \ell_\beta \gamma \right) = \frac{e^2 \, \left( m_\alpha^2 - m_\beta^2 \right)^3}{16 \, \pi \, m_\alpha}  \left[ \left|\left(K_2^L\right)^{\beta\alpha}\right|^2 + \left|\left(K_2^R\right)^{\beta \alpha}\right|^2 \right] \, .
\end{equation*}
}
Then, we can compare our analytical results with
  those found in~\cite{Bauer:2019gfk}. Assuming that the only
  non-vanishing couplings are the ones involving the $\mu \mu$ and $e
  \mu$ flavor combinations, and making the replacements
\begin{equation}
  S^{\mu \mu} = i \frac{ m_\mu \, c_{\mu \mu}}{f} \quad , \quad S_L^{e \mu} = i \frac{m_\mu \, \left(k_e\right)_{e \mu}}{f} \quad , \quad S_R^{e \mu} = i \frac{m_\mu \, \left(k_E\right)_{e \mu}}{f} \, ,
\end{equation}
full agreement is recovered.

\subsection[\texorpdfstring{$\ell_\alpha^- \to \ell_\beta^- \ell_\beta^- \ell_\beta^+$}{\unichar{"2113}\unichar{"03B1}- \unichar{"2192} \unichar{"2113}\unichar{"03B2}- \unichar{"2113}\unichar{"03B2}- \unichar{"2113}\unichar{"03B2}+}]{$\boldsymbol{\ell_\alpha^- \to \ell_\beta^- \ell_\beta^- \ell_\beta^+}$}

\begin{figure}[!t]
  \centering
  \includegraphics[width=0.82\linewidth]{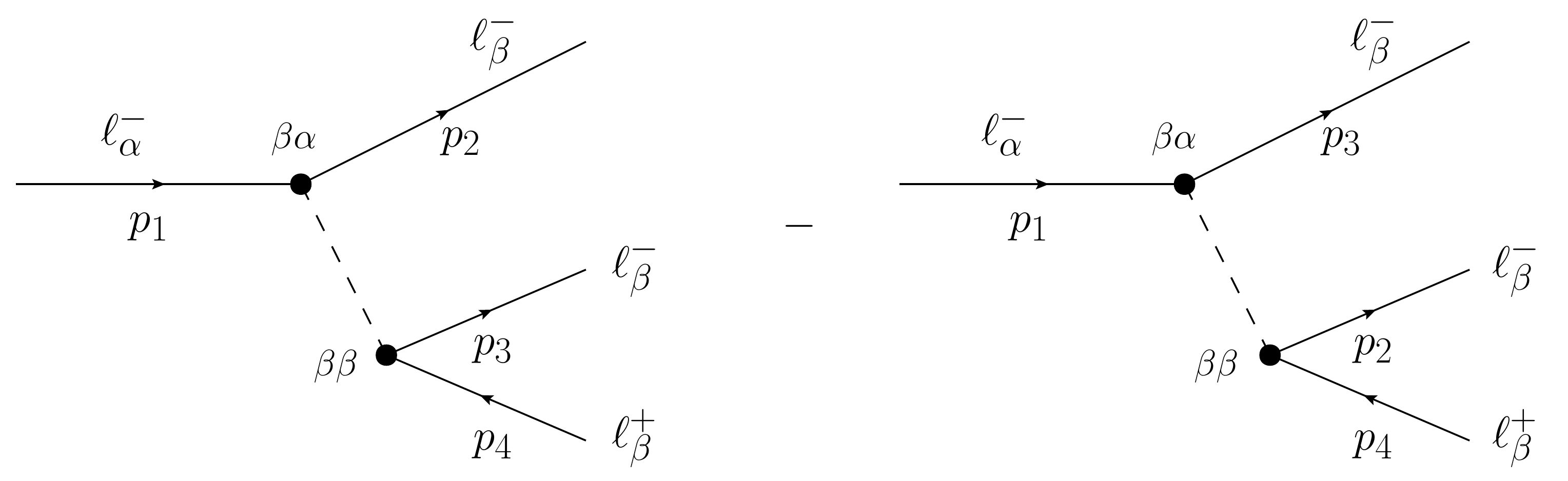}
  \caption{Tree-level Feynman diagrams contributing to the process $\ell_{\alpha}^{-} \to \ell_{\beta}^{-} \ell_{\beta}^- \ell_{\beta}^+$ described by the effective Lagrangian in Eq.~\eqref{eq:lagS}.}
  \label{fig:Diagrams}
\end{figure}

Complete expressions for the $\ell_{\alpha}^{-} \to \ell_{\beta}^{-}
\ell_{\beta}^- \ell_{\beta}^+$ decay width in the absence of $\phi$
can be found in~\cite{Abada:2014kba}. Here we are interested in the
new contributions mediated by the scalar $\phi$, which are given by
the Feynman diagrams shown in Fig.~\ref{fig:Diagrams}. It is
straightforward to derive the associated amplitude, given by
\begin{equation}
  \begin{split}
    \mathcal{M}_\phi &= \bar{u} \left( p_3 \right) i \left( S^{\beta \beta} P_L + S^{\beta \beta *} P_R \right) v \left(p_4\right) \frac{i}{q^2 + i \varepsilon} \, \bar{u} \left(p_2\right) i \left( S_L^{\beta \alpha} P_L + S_R^{\beta \alpha} P_R \right) u \left( p_1 \right) \\
    & - \bar{u} \left( p_2 \right) i \left( S^{\beta \beta} P_L + S^{\beta \beta *} P_R \right) v \left(p_4\right) \frac{i}{k^2 + i \varepsilon} \, \bar{u} \left(p_3\right) i \left( S_L^{\beta \alpha} P_L + S_R^{\beta \alpha} P_R \right) u \left( p_1 \right) \, .
  \end{split}
  \label{eq:AmplitudeS}
\end{equation}
Here $q = p_1 - p_2$ and $k = p_1 - p_3$ are
the $\phi$ virtual momenta and we have explicitly indicated the flavor
indices of the $S_{L,R}$ coefficients. The total decay width can then
be written as
\begin{equation}
  \Gamma\left( \ell_{\alpha}^{-} \to \ell_{\beta}^{-}
\ell_{\beta}^- \ell_{\beta}^+ \right) = \Gamma_{\bar{\phi}}\left( \ell_{\alpha}^{-} \to \ell_{\beta}^{-}
\ell_{\beta}^- \ell_{\beta}^+ \right) + \Gamma_{\phi}\left( \ell_{\alpha}^{-} \to \ell_{\beta}^{-}
\ell_{\beta}^- \ell_{\beta}^+ \right) \, ,
\end{equation}
where $\Gamma_{\bar{\phi}}$ is the decay width in the absence of
$\phi$, given in~\cite{Abada:2014kba}, and
\begin{equation}
	\begin{split}
		& \Gamma_\phi\left( \ell_{\alpha}^{-} \to \ell_{\beta}^{-}
\ell_{\beta}^- \ell_{\beta}^+ \right) = \\
		& \frac{m_{\alpha}}{512 \pi^{3}} \Biggl\{ \left( \left| S_{L}^{\beta \alpha} \right|^{2} + \left| S_{R}^{\beta \alpha} \right|^{2} \right) \left\{ \left| S^{\beta \beta} \right|^{2} \left( 4 \log \frac{m_{\alpha}}{m_{\beta}} - \frac{49}{6} \right) - \frac{2}{6} \left[ \left( S^{\beta \beta *} \right)^2 + \left( S^{\beta \beta} \right)^2 \right] \right\} \Biggr. \\
		& - \frac{m_\alpha^2}{6} \biggl\{ S^{\beta \alpha}_L S^{\beta \beta} A^{S *}_{LL} +2 S_{L}^{\beta \alpha} S^{\beta \beta *} A_{L R}^{S *}+2 S_{R}^{\beta \alpha} S^{\beta \beta} A_{R L}^{S *}+S_{R}^{\beta \alpha} S^{\beta \beta *} A_{R R}^{S *}\biggr. \\
		& - 12 \left( S^{\beta \alpha}_L S^{\beta \beta} A^{T *}_{LL} + S^{\beta \alpha}_R S^{\beta \beta *} A^{T *}_ {RR} \right) - 4 \left( S^{\beta \alpha}_R S^{\beta \beta} A^{V *}_{RL} + S^{\beta \alpha}_L S^{\beta \beta *} A^{V *}_{LR} \right) \\
		& \Biggl. \biggl. + \ 6 e^2 \left[ S^{\beta \alpha}_R S^{\beta \beta} \left(K_2^{L}\right)^{\beta \alpha \ast} + S^{\beta \alpha}_L S^{\beta \beta *} \left(K_2^{R}\right)^{\beta \alpha \ast} \right] + \cc \biggr\} \Biggr\} \, ,
	\end{split}
	\label{eq:widthS}
\end{equation}
where in this expression $A_{X Y}^I = \left( A_{X Y}^I \right)^{\beta \beta \beta \alpha}$. In writing Eq.~\eqref{eq:widthS} we have only kept the lowest order
terms in powers of $m_\beta$ for each possible combination of
couplings. This is equivalent to 0th order for all terms, with the
exception of the ones in the first line, where the factor $\log
\frac{m_\alpha}{m_\beta}$ avoids the appearance of an infrared
divergence. An expression including terms up to first order in
$m_\beta$ is given in Appendix~\ref{sec:polar}.

\subsection[\texorpdfstring{$\ell_\alpha^- \to \ell_\beta^- \ell_\gamma^- \ell_\gamma^+$}{\unichar{"2113}\unichar{"03B1}- \unichar{"2192} \unichar{"2113}\unichar{"03B2}- \unichar{"2113}\unichar{"03B3}- \unichar{"2113}\unichar{"03B3}+}]{$\boldsymbol{\ell_\alpha^- \to \ell_\beta^- \ell_\gamma^- \ell_\gamma^+}$}

\begin{figure}[!t]
  \centering
  \includegraphics[width=0.82\linewidth]{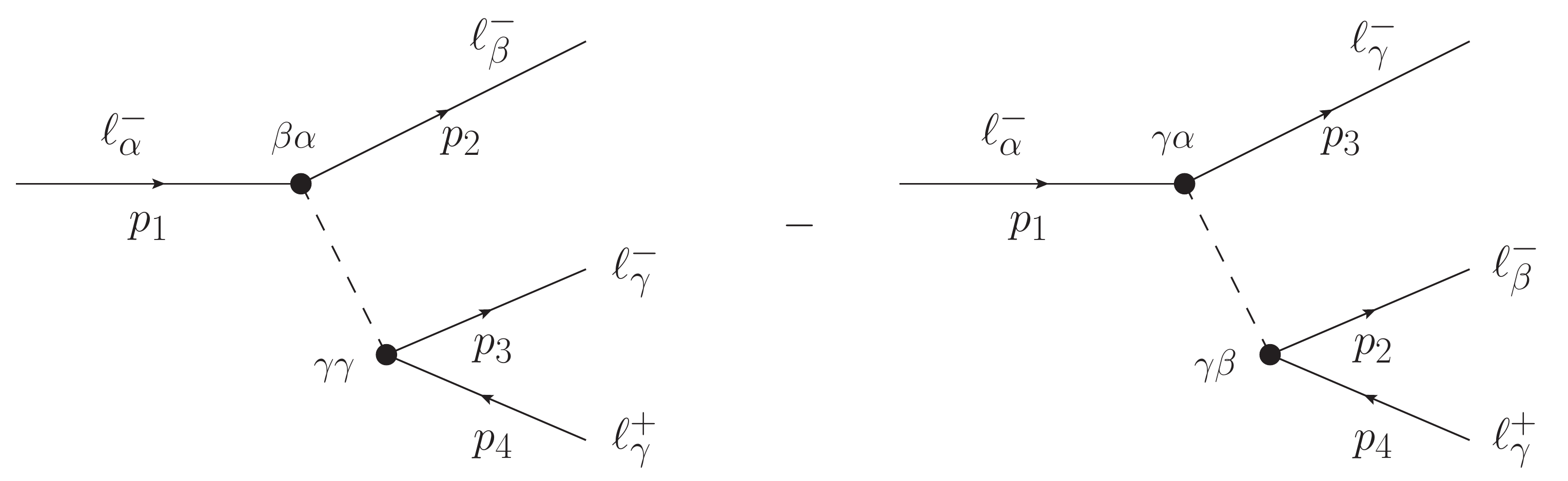}
  \caption{Tree-level Feynman diagrams contributing to the process $\ell_{\alpha}^{-} \to \ell_{\beta}^{-} \ell_{\gamma}^- \ell_{\gamma}^+$ described by the effective Lagrangian in Eq.~\eqref{eq:lagS}.}
  \label{fig:Diagramsbeta-2gamma}
\end{figure}

Again, complete expressions for the $\ell_{\alpha}^{-} \to
\ell_{\beta}^{-} \ell_{\gamma}^- \ell_{\gamma}^+$ decay width in the
absence of $\phi$ can be found in~\cite{Abada:2014kba}. The new
contributions mediated by the scalar $\phi$ are obtained from the
Feynman diagrams shown in Fig.~\ref{fig:Diagramsbeta-2gamma}. While
the diagram on the left involves a flavor conserving ($\gamma \gamma$)
and a flavor violating ($\beta \alpha$) vertex, both vertices in the
diagram on the right violate flavor ($\gamma \alpha$ and $\gamma
\beta$). The associated amplitude is slightly different from that of
the previous process and is given by
\begin{equation}
  \begin{split}
    \mathcal{M}_\phi &= \bar{u} \left( p_3 \right) i \left( S^{\gamma \gamma} P_L + S^{\gamma \gamma *} P_R \right) v \left(p_4\right) \frac{i}{q^2 + i \varepsilon} \, \bar{u} \left(p_2\right) i \left( S_L^{\beta \alpha} P_L + S_R^{\beta \alpha} P_R \right) u \left( p_1 \right) \\
    & - \bar{u} \left( p_2 \right) i \left( S_L^{\gamma \beta} P_L + S_R^{\gamma \beta} P_R \right) v \left(p_4\right) \frac{i}{k^2 + i \varepsilon} \, \bar{u} \left(p_3\right) i \left( S_L^{\gamma \alpha} P_L + S_R^{\gamma \alpha} P_R \right) u \left( p_1 \right) \, .
  \end{split}
  \label{eq:AmplitudeSbeta-2gamma}
\end{equation}
Finally, the total decay width can be written as
\begin{equation}
  \Gamma\left( \ell_{\alpha}^{-} \to \ell_{\beta}^{-}
\ell_{\gamma}^- \ell_{\gamma}^+ \right) = \Gamma_{\bar{\phi}}\left( \ell_{\alpha}^{-} \to \ell_{\beta}^{-}
\ell_{\gamma}^- \ell_{\gamma}^+ \right) + \Gamma_{\phi}\left( \ell_{\alpha}^{-} \to \ell_{\beta}^{-}
\ell_{\gamma}^- \ell_{\gamma}^+ \right) \, ,
\end{equation}
where $\Gamma_{\bar{\phi}}$ is the decay width in the absence of
$\phi$, given in~\cite{Abada:2014kba}, and
\begin{equation}
	\begin{split}
		& \Gamma_\phi\left( \ell_{\alpha}^{-} \to \ell_{\beta}^{-}
\ell_{\gamma}^- \ell_{\gamma}^+ \right) = \\
		& \frac{m_{\alpha}}{512 \pi^{3}} \Biggl\{ \left( \left| S_{L}^{\beta \alpha} \right|^{2} + \left| S_{R}^{\beta \alpha} \right|^{2} \right) \left\{ \left| S^{\gamma \gamma} \right|^{2} \left( 4 \log \frac{m_{\alpha}}{m_{\gamma}} - \frac{23}{3} \right) - \frac{1}{3} \left[ \left( S^{\gamma \gamma *} \right)^2 + \left( S^{\gamma \gamma} \right)^2 \right] \right\} \Biggr. \\
		& + \left( \left| S_{L}^{\gamma \alpha} \right|^{2} + \left| S_{R}^{\gamma \alpha} \right|^{2} \right) \left( \left| S_{L}^{\gamma \beta} \right|^{2} + \left| S_{R}^{\gamma \beta} \right|^{2} \right) \left( 2 \log \frac{m_{\alpha}}{m^{\max}_f} - 3 \right) \\
		& - \frac{1}{2} \left[ S^{\gamma \gamma} \left( S_L^{\beta \alpha} S_L^{\gamma \alpha *} S_L^{\gamma \beta *} + S_R^{\beta \alpha *} S_R^{\gamma \alpha} S_R^{\gamma \beta} \right)  + \cc \right] \\
		& + \frac{m_\alpha^2}{6} \biggl\{ S_L^{\gamma \alpha} S_L^{\gamma \beta} A_{LL}^{S *} + S_R^{\gamma \alpha} S_R^{\gamma \beta} A_{RR}^{S *} - 2 S^{\gamma \gamma} \left( S_L^{\beta \alpha} A_{LL}^{S *} + S_L^{\beta \alpha *} A_{LR}^{S} + S_R^{\beta \alpha} A_{RL}^{S *} + S_R^{\beta \alpha *} A_{RR}^{S}  \right)  \biggr. \\
		& + 4 \left( S_L^{\gamma \alpha} S_R^{\gamma \beta} A_{LR}^{V *} + S_R^{\gamma \alpha} S_L^{\gamma \beta} A_{RL}^{V *} \right) + 12 \left( S_L^{\gamma \alpha} S_L^{\gamma \beta} A_{LL}^{T *} + S_R^{\gamma \alpha} S_R^{\gamma \beta} A_{RR}^{T *} \right) \\
		& \Biggl. \biggl. - 6 e^2 \left[ S_L^{\gamma \alpha} S_R^{\gamma \beta} \left(K_2^{R}\right)^{\beta \alpha \ast} + S_R^{\gamma \alpha} S_L^{\gamma \beta} \left(K_2^{L}\right)^{\beta \alpha \ast} \right] + \cc \biggr\} \Biggr\},
	\end{split}
	\label{eq:widthSbeta-2gamma}
\end{equation}
where in this expression $A_{X Y}^I = \left( A_{X Y}^I \right)^{\gamma \gamma \beta \alpha}$. Also here $m^{\max}_f = \text{max}\left(m_\beta, m_\gamma\right)$ and then
the expression depends on the process in question. Once again, we have
only kept the lowest order terms in powers of $m_\beta$ and $m_\gamma$
for each possible combination of couplings.

\begin{figure}[!t]
  \centering
  \includegraphics[width=0.82\linewidth]{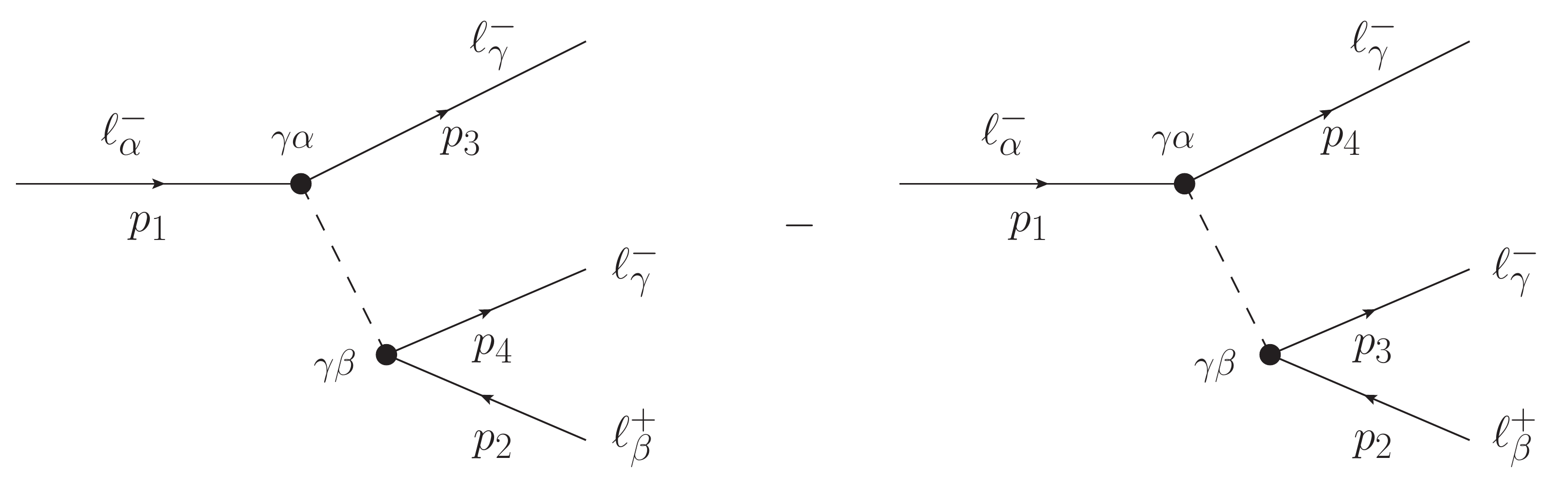}
  \caption{Tree-level Feynman diagrams contributing to the process $\ell_{\alpha}^{-} \to \ell_{\beta}^{+} \ell_{\gamma}^- \ell_{\gamma}^-$ described by the effective Lagrangian in Eq.~\eqref{eq:lagS}.}
  \label{fig:Diagramsbeta+2gamma}
\end{figure}

\subsection[\texorpdfstring{$\ell_\alpha^- \to \ell_\beta^+ \ell_\gamma^- \ell_\gamma^-$}{\unichar{"2113}\unichar{"03B1}- \unichar{"2192} \unichar{"2113}\unichar{"03B2}+ \unichar{"2113}\unichar{"03B3}- \unichar{"2113}\unichar{"03B3}-}]{$\boldsymbol{\ell_\alpha^- \to \ell_\beta^+ \ell_\gamma^- \ell_\gamma^-}$}

Also for this process, complete expressions for the $\ell_{\alpha}^{-}
\to \ell_{\beta}^{+} \ell_{\gamma}^- \ell_{\gamma}^-$ decay width in
the absence of $\phi$ can be found in~\cite{Abada:2014kba}. The new
contributions mediated by the scalar $\phi$ are given by the Feynman
diagrams shown in Fig.~\ref{fig:Diagramsbeta+2gamma}. We note that
both vertices are necessarily flavor violating. The associated
amplitude is given in this case by
\begin{equation}
  \begin{split}
    \mathcal{M}_\phi &= \bar{u} \left( p_4 \right) i \left( S_L^{\gamma \beta} P_L + S_R^{\gamma \beta} P_R \right) v \left(p_2\right) \frac{i}{q^2 + i \varepsilon} \, \bar{u} \left(p_3\right) i \left( S_L^{\gamma \alpha} P_L + S_R^{\gamma \alpha} P_R \right) u \left( p_1 \right) \\
    & - \bar{u} \left( p_3 \right) i \left( S_L^{\gamma \beta} P_L + S_R^{\gamma \beta} P_R \right) v \left(p_2\right) \frac{i}{k^2 + i \varepsilon} \, \bar{u} \left(p_4\right) i \left( S_L^{\gamma \alpha} P_L + S_R^{\gamma \alpha} P_R \right) u \left( p_1 \right) \, .
  \end{split}
  \label{eq:AmplitudeSbeta+2gamma}
\end{equation}
Here $q = p_1 - p_3$ and $k = p_1 - p_4$ are different from their
definitions in the processes above. Writing one more time the decay
width as the sum of two contributions,
\begin{equation}
  \Gamma\left( \ell_{\alpha}^{-} \to \ell_{\beta}^{+}
\ell_{\gamma}^- \ell_{\gamma}^- \right) = \Gamma_{\bar{\phi}}\left( \ell_{\alpha}^{-} \to \ell_{\beta}^{+}
\ell_{\gamma}^- \ell_{\gamma}^- \right) + \Gamma_{\phi}\left( \ell_{\alpha}^{-} \to \ell_{\beta}^{+}
\ell_{\gamma}^- \ell_{\gamma}^- \right) \, ,
\end{equation}
where $\Gamma_{\bar{\phi}}$ is the decay width in the absence of $\phi$, given in~\cite{Abada:2014kba}, we find that
\begin{equation}
	\begin{split}
		& \Gamma_\phi\left( \ell_{\alpha}^{-} \to \ell_{\beta}^{+}
\ell_{\gamma}^- \ell_{\gamma}^- \right) = \\
		& \frac{m_{\alpha}}{512 \pi^{3}} \left\{ \left( \left| S_{L}^{\gamma \alpha} \right|^{2} + \left| S_{R}^{\gamma \alpha} \right|^{2} \right) \left( \left| S_{L}^{\gamma \beta} \right|^{2} + \left| S_{R}^{\gamma \beta} \right|^{2} \right) \left( 2 \log \frac{m_{\alpha}}{m^{\max}_f} - 3 \right) \right. \\
		& - \frac{1}{2} \left( \left| S_{L}^{\gamma \alpha} \right|^{2} \left| S_{L}^{\gamma \beta} \right|^{2} + \left| S_{R}^{\gamma \alpha} \right|^{2} \left| S_{R}^{\gamma \beta} \right|^{2} \right) \\
		& + \frac{m_\alpha^2}{6} \biggl[ - S_L^{\gamma \alpha} S_L^{\gamma \beta} A_{LL}^{S *} - S_R^{\gamma \alpha} S_R^{\gamma \beta} A_{RR}^{S *} - 2 \left( S_L^{\gamma \alpha} S_R^{\gamma \beta} A_{RL}^{S *} + S_R^{\gamma \alpha} S_L^{\gamma \beta} A_{LR}^{S *} \right) \biggr. \\
		& \Biggl. \biggl. + 4 \left( S_L^{\gamma \alpha} S_R^{\gamma \beta} A_{RL}^{V *} + S_R^{\gamma \alpha} S_L^{\gamma \beta} A_{LR}^{V *} \right) + 12 \left( S_L^{\gamma \alpha} S_L^{\gamma \beta} A_{LL}^{T *} + S_R^{\gamma \alpha} S_R^{\gamma \beta} A_{RR}^{T *} \right) + \cc \biggr] \Biggr\},
	\end{split}
	\label{eq:widthSbeta+2gamma}
\end{equation}
where in this expression $A_{X Y}^I = \left( A_{X Y}^I \right)^{\gamma \beta \gamma \alpha}$ and $m^{\max}_f = \text{max}\left(m_\beta, m_\gamma\right)$.

\subsection{Lepton magnetic and electric dipole moments} \label{sec:AMM_EDM}

\begin{figure}[tb]
  \centering
  \includegraphics[width=0.6\linewidth]{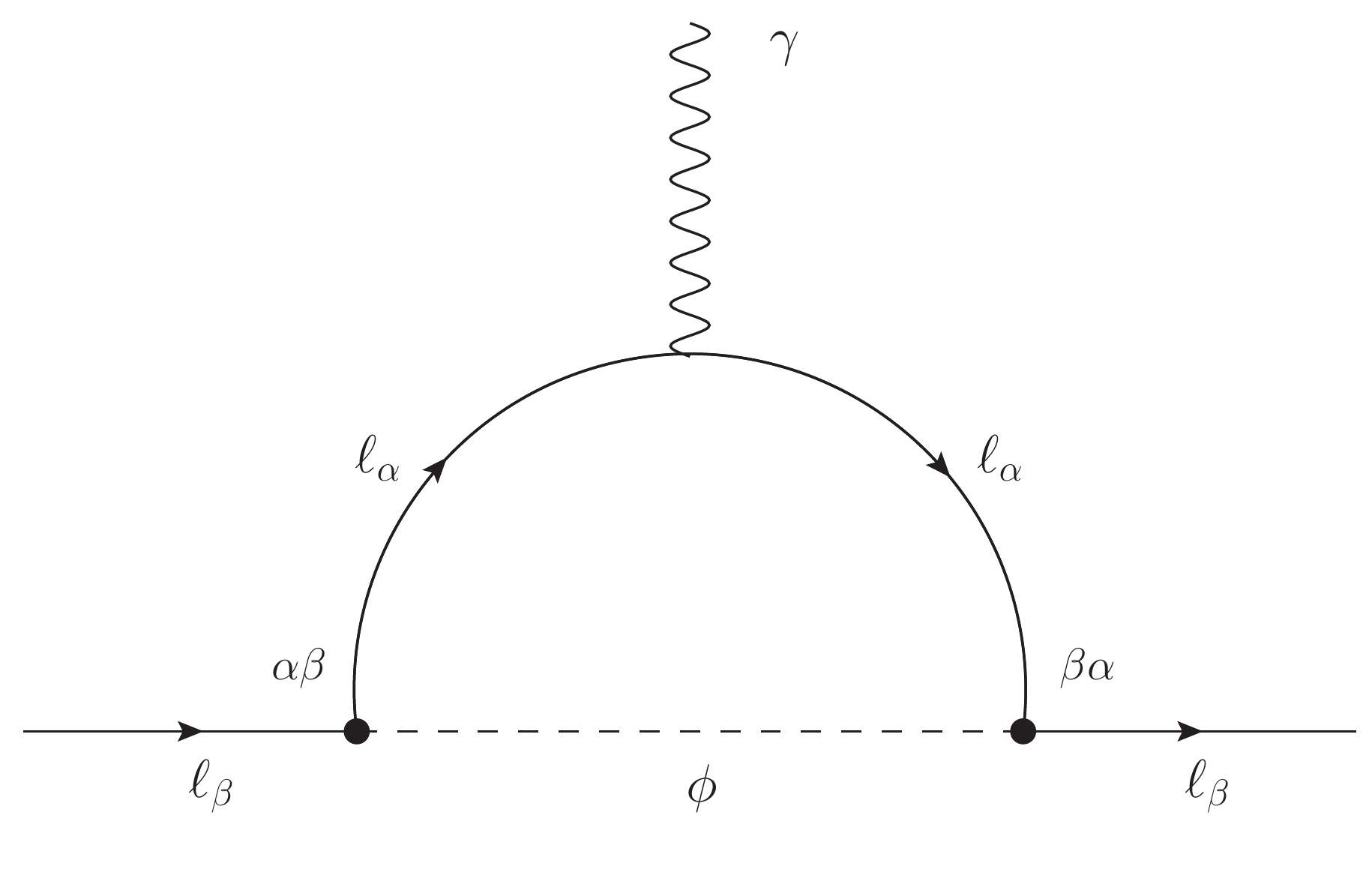}
  \caption{Feynman diagram for the one-loop contribution to the
    anomalous magnetic moment of charged leptons given by the
    interaction in Eq.~\eqref{eq:lagS}. The flavor
      indices of the couplings contributing to the diagram have been
      specified in the vertices. The relation between these couplings,
      which we generically denote as $C_{L,R}$ and
      $\widetilde{C}_{L,R}$, and the $S_{L,R}$ couplings in the
      effective Lagrangian of Eq.~\eqref{eq:lagS} depends on the
      flavor states involved in the diagram. See text for a detailed
      explanation.}
    \label{fig:one-loop_g-2}
\end{figure}

We finally consider the magnetic and electric dipole
  moments of the charged leptons. These can be described by the
  effective Lagrangians
\begin{align}
  \mathcal{L}_{\rm AMM} &= \frac{e}{2 \, m_\alpha } a_{\alpha} \, \overline{\ell}_\alpha \, \sigma^{\mu \nu} F_{\mu \nu} \, \ell_\alpha \, , \label{eq:eff_AMM} \\
  \mathcal{L}_{\rm EDM} &= - \frac{i}{2} d_{\alpha} \, \overline{\ell}_\alpha \, \sigma^{\mu \nu} F_{\mu \nu} \, \gamma_5 \, \ell_\alpha \, . \label{eq:eff_EDM}
\end{align}
The charged lepton dipole moments receive contributions mediated by
the scalar $\phi$, as shown in
Fig.~\ref{fig:one-loop_g-2}.~\footnote{Two-loop Barr-Zee
  contributions~\cite{Barr:1990vd} to the charged leptons AMMs and EDMs
  can also be considered. However, while these might be relevant in
  some cases, we will assume that the $S_{L,R}$ couplings can at most
  have mild hierarchies among different flavors, hence making them
  numerically irrelevant with respect to the one-loop contributions
  considered here.} In the following we denote the chiral couplings in
the $\bar{\ell}_\alpha - \ell_\beta - \phi$ vertex as $C_L^{\alpha
  \beta}$ and $C_R^{\alpha \beta}$, whereas the chiral couplings in
the $\bar{\ell}_\beta - \ell_\alpha - \phi$ vertex are denoted as
$\widetilde{C}_L^{\beta \alpha}$ and $\widetilde{C}_R^{\beta
  \alpha}$. The $C_{L,R}$ and $\widetilde{C}_{L,R}$ couplings are
obviously related to the $S_{L,R}$ couplings in the effective
Lagrangian of Eq.~\eqref{eq:lagS}, but this relation depends on the
flavor states involved in the diagram, as discussed below. The
amplitude associated to the diagram in Fig.~\ref{fig:one-loop_g-2} can
be written as
\begin{align}
  i \mathcal{M} = & \int \frac{\text{d}^4 q}{\left( 2 \pi\right)^4} \overline{u}_\ell \left(p^\prime, m_\alpha \right) \left[ i \left( C_L^{\alpha \beta} \, P_L + C_R^{\alpha \beta} \, P_R \right) \right] \frac{i \left( \slashed{p}^\prime + \slashed{q} + m_\beta \right)}{\left( p^\prime + q \right)^2 - m_\beta^{2}} \left( - i \, e \, \gamma^\mu \right) \frac{i \left( \slashed{p} + \slashed{q} + m_\beta \right)}{\left( p + q \right)^2 - m_\beta^{2}} \nonumber \\
  & \left[ i \left( \widetilde{C}_L^{\beta \alpha} \, P_L + \widetilde{C}_R^{\beta \alpha} \, P_R \right) \right] \frac{i}{q^2} u_\ell \left( p, m_\alpha \right) \, \varepsilon_\mu^* \left( k \right),
\end{align}
where $m_\alpha$ and $m_\beta$ are the masses of the external and
internal leptons, respectively, and we sum over the index $\beta$. One
must now compare to the equivalent amplitude obtained with the
effective Lagrangians in Eqs.~\eqref{eq:eff_AMM} and
\eqref{eq:eff_EDM}. After some algebra, one finds that the scalar
$\phi$ induces the contributions to the anomalous magnetic moments
(AMMs) and electric dipole moments (EDMs) of the charged leptons
\begin{align}
  \Delta a_{\alpha} = &  \frac{1}{32 \, \pi^2 \, m_\alpha^4} \left\{2 \, m_\alpha \, m_\beta \left[m_\alpha^2 + \left( m_\alpha^2 - m_\beta^2 \right) \log \frac{m_\beta^{2}}{ \left| m_\beta^{2} - m_\alpha^2 \right| } \right] \left(C_L^{\alpha \beta} \, \widetilde{C}_L^{\beta \alpha} + C_R^{\alpha \beta} \, \widetilde{C}_R^{\beta \alpha} \right) \right. \nonumber \\
  & \left. - \left[ m_\alpha^2 \left(m_\alpha^2 - 2 \, m_\beta^2 \right) - 2 \, m_\beta^2 \left(m_\alpha^2 - m_\beta^2 \right) \log \frac{m_\beta^{2}}{ \left| m_\beta^{2} - m_\alpha^2 \right| } \right] \left( C_L^{\alpha \beta} \, \widetilde{C}_R^{\beta \alpha} + C_R^{\alpha \beta} \, \widetilde{C}_L^{\beta \alpha} \right) \right\} \, ,
\label{eq:general_a}
\end{align}
and
\begin{equation}
  d_{\alpha} = \frac{ i \, e \, m_\beta}{32 \, \pi^2 \, m_\alpha^4} \left( C_L^{\alpha \beta} \, \widetilde{C}_L^{\beta \alpha} - C_R^{\alpha \beta} \, \widetilde{C}_R^{\beta \alpha} \right) \left[ m_\alpha^2 + \left(m_\alpha^2 - m_\beta^{2} \right) \log \frac{m_\beta^{2}}{\left| m_\beta^{2} - m_\alpha^2 \right|} \right] \, .
\label{eq:general_d}
\end{equation}
These analytical results have been checked with the
help of {\tt Package-X}~\cite{Patel:2015tea}. We note again that a
sum over the index $\beta$ is performed in
Eqs.~\eqref{eq:general_a} and \eqref{eq:general_d}. Therefore, they
include both flavor diagonal as well as flavor off-diagonal
contributions to the dipole moments. We now consider these
contributions separately and study their behavior in specific
limits:
\begin{enumerate}

\item Flavor off-diagonal contribution with $m_\beta \ll m_\alpha$

In this case the $C_{L,R}$ and $\widetilde{C}_{L,R}$
couplings are related to the $S_{L,R}$ couplings in
Eq.~\eqref{eq:lagS} as $\widetilde{C}_{L,R}^{\beta \alpha} =
S_{L,R}^{\beta \alpha}$ and $C_{L,R}^{\alpha \beta} =
S_{R,L}^{\beta \alpha *}$ and the expressions simplify to
\begin{align}
  \Delta a_{\alpha} = \, & \frac{1}{32 \, \pi^2 \, m_\alpha} \left[ - m_\alpha \left( \left| S_{L}^{\beta \alpha} \right|^2 + \left| S_{R}^{\beta \alpha} \right|^2 \right) \right.  \nonumber \\
  & \left. + \, 4 \, m_\beta \, \text{Re} \left( S_{R}^{\beta \alpha} \, S_{L}^{\beta \alpha *} \right) \left( 1 + \log \frac{m_\beta^{2}}{m_\alpha^2} \right) \right] + \mathcal{O} \left( m_\beta^{2} \right),
\end{align}
and
\begin{equation}
  d_{\alpha} = \frac{e \, m_\beta}{16 \, \pi^2 \, m_\alpha^2} \, \text{Im} \left( S_{R}^{\beta \alpha} \, S_{L}^{\beta \alpha *} \right) \left( 1 + \log \frac{m_\beta^{2}}{m_\alpha^2} \right)  + \mathcal{O} \left( m_\beta^{3} \right).
\end{equation}

\item Flavor off-diagonal contribution with $m_\beta \gg m_\alpha$

In this case the generic $C_{L,R}$ and
$\widetilde{C}_{L,R}$ couplings are related to the $S_{L,R}$
couplings as $\widetilde{C}_{L,R}^{\beta \alpha} = S_{R,L}^{\alpha
\beta *}$ and $C_{L,R}^{\alpha \beta} = S_{L,R}^{\alpha \beta}$,
giving us
\begin{align}
  \Delta a_{\alpha} = \, & \frac{m_\alpha}{16 \, \pi^2 \, m_\beta} \, \left[ \text{Re} \left( S_{R}^{\alpha \beta} \, S_{L}^{\alpha \beta *} \right) + \frac{m_\alpha}{6 \, m_\beta} \left( \left| S_{L}^{\beta \alpha} \right|^2 + \left| S_{R}^{\beta \alpha} \right|^2 \right) \right] \color{black} + \mathcal{O} \left( m_\alpha^{3} \right),
\end{align}
and
\begin{equation}
  d_{\alpha} = \frac{e}{32 \, \pi^2 \, m_\beta} \, \text{Im} \left( S_{R}^{\alpha \beta} \, S_{L}^{\alpha \beta *} \right) + \mathcal{O} \left( m_\alpha^{2} \right).
\end{equation}

\item Flavor diagonal contribution, i.e. $m_\beta = m_\alpha$

Finally, in this case we have $C_L^{\alpha \alpha} = \widetilde{C}_L^{\alpha \alpha} = S^{\alpha \alpha}$ and $C_R^{\alpha \alpha} = \widetilde{C}_R^{\alpha \alpha} = S^{\alpha \alpha *}$, and we find the simple expression
\begin{equation} \label{eq:anomalous_magnetic_moment_diag}
  \Delta a_{\alpha} = \frac{1}{16 \pi^{2}} \left[ 3 \, \left( \text{Re} \, S^{\alpha \alpha} \right)^2 - \left( \text{Im} \, S^{\alpha \alpha} \right)^2 \right] 
\end{equation}
for the AMM of the charged lepton $\ell_\alpha$, which agrees with previous results in the literature. In particular,
it matches exactly the expression given in~\cite{Botella:2020xzf} in
the limit of a massless scalar, with the equivalence
\begin{equation}
  - \frac{m_\alpha}{v} \, a_\alpha^S = \frac{1}{2} \left( S^{\alpha \alpha} + S^{\alpha \alpha *} \right) \, , \quad - i \, \frac{m_\alpha}{v} \, b_\alpha^S = - \frac{1}{2} \left( S^{\alpha \alpha} - S^{\alpha \alpha *} \right) \, .
\end{equation}
Regarding the expression for the EDM, it also acquires a very simple
form in this case,
\begin{equation} \label{eq:EDM_diag}
  d_{\alpha} = - \frac{e}{8 \, \pi^2 \, m_\alpha} \left( \text{Re} \, S^{\alpha \alpha} \right) \, \left( \text{Im} \, S^{\alpha \alpha} \right) \, .
\end{equation}
This expression agrees with the one given
in~\cite{PhysRevD.93.035006} just by identifying $\text{Re} \,
S^{\alpha \alpha} = - \lambda_S^\ell$ and $\text{Im} \, S^{\alpha 
\alpha} = \lambda_P^\ell$. Notice that
Eqs.~\eqref{eq:anomalous_magnetic_moment_diag} and
\eqref{eq:EDM_diag} are both exact results for the diagonal
contributions to the dipole moments.

\end{enumerate}

\section{Phenomenological discussion}
\label{sec:pheno}

After deriving analytical expressions for several
  leptonic observables of interest we now discuss their associated
  phenomenology.

\subsection[Searches for \texorpdfstring{$\ell_\alpha \to \ell_\beta \, \phi$}{\unichar{"2113}\unichar{"03B1} \unichar{"2192} \unichar{"2113}\unichar{"03B2} \unichar{"03D5}}]{Searches for $\boldsymbol{\ell_\alpha \to \ell_\beta \, \phi}$}
\label{sec:phenobetaphi}

Several searches for $\ell_\alpha \to \ell_\beta \, \phi$ have been
performed and used to set experimental contraints on the off-diagonal
$S_A^{\beta \alpha}$ effective couplings. Let us start with muon
decays. The strongest limit on the branching ratio for the 2-body
decay $\mu^+ \to e^+ \, \phi$ was obtained at TRIUMF, finding $\BR
\left(\mu \to e \, \phi\right) < 2.6 \times
10^{-6}$ at 90\% C.L.~\cite{Jodidio:1986mz}. However, as explained
in~\cite{Hirsch:2009ee}, this experimental limit must be applied with
care to the general scenario considered here. The reason is that the
experimental setup in \cite{Jodidio:1986mz} uses a muon beam that is
highly polarized in the direction opposite to the muon momentum and
concentrates the search in the forward region. This reduces the
background from the SM process $\mu^+ \to e^+ \nu_e \, \bar \nu_\mu$,
which is strongly suppressed in this region, but also reduces the
$\mu^+ \to e^+ \, \phi$ signal unless the $\phi-e-\mu$ coupling is
purely right-handed. Therefore, we obtain a limit valid only when
$S_L^{e \mu} = 0$:
\begin{equation}
  S_L^{e \mu} = 0 \quad \Rightarrow \quad \left| S^{e \mu}_R \right| < 2.7 \times 10^{-11} \, .
\end{equation}
A more general limit can also be derived from
~\cite{Jodidio:1986mz}. Using the spin processed data shown in Fig.(7)
of~\cite{Jodidio:1986mz}, the authors of~\cite{Hirsch:2009ee} obtained
the conservative bound $\BR \left( \mu \to e \, \phi \right) \lesssim
10^{-5}$, valid for any chiral structure of the $S_A^{e \mu}$
couplings. This bound is similar to the more recent limit obtained by
the TWIST collaboration~\cite{Bayes:2014lxz}, also in the $\sim
10^{-5}$ ballpark. With this value, one finds an upper limit on the
$e-\mu$ flavor violating couplings of~\footnote{See also the recent
  \cite{Calibbi:2020jvd} for a comprehensive discussion of the
  experimental limit of \cite{Jodidio:1986mz} and how this gets
  altered for different chiral structures of the $S_A^{e \mu}$
  couplings.}
\begin{equation} \label{eq:limemu1}
  \left| S^{e \mu} \right| < 5.3 \times 10^{-11} \, .
\end{equation}
where we have defined the convenient combination
\begin{equation}
  \left| S^{\beta \alpha} \right| = \left( \left| S^{\beta \alpha}_L \right|^2 + \left| S^{\beta \alpha}_R \right|^2 \right)^{1/2} \, .
\end{equation}

Several strategies can be followed for newer $\mu \to e \, \phi$
searches. The authors of \cite{Calibbi:2020jvd} advocate for a new
phase of the MEG-II experiment, reconfigured to search for $\mu \to e
\, \phi$ by placing a Lyso calorimeter in the forward direction. Also,
as pointed out in~\cite{Perrevoort:2018ttp,Perrevoortthesis} and
recently discussed in~\cite{Calibbi:2020jvd} as well, the limit in
Eq.~\eqref{eq:limemu1} can be substantially improved by the Mu3e
experiment by looking for a bump in the continuous Michel
spectrum. The detailed analysis in \cite{Perrevoortthesis} shows that
$\mu \to e \, \phi$ branching ratios above $7.3 \times 10^{-8}$ can be
ruled out at 90\% C.L.. This would imply a sensitivity to an $\left|
S^{e \mu} \right|$ effective coupling as low as $4.5 \times 10^{-12}$,
improving an order of magnitude with respect to the limit in
Eq.~\eqref{eq:limemu1}.

Turning to $\tau$ decays, the currently best experimental limits were
set by the ARGUS collaboration~\cite{Albrecht:1995ht}, which found
\begin{equation}
  \begin{split}
    & \frac{\BR\left(\tau \to e \, \phi \, \right)}{\BR \left(\tau \to e \, \nu \, \bar{\nu} \right) } < 0.015 \, , \\
    & \frac{\BR\left(\tau \to \mu \, \phi \right)}{\BR \left(\tau \to \mu \, \nu \, \bar{\nu} \right) } < 0.026 \, ,
  \end{split}
\end{equation}
at 95\% C.L.. These limits are weaker than those for muon decays, but
still lead to stringent constraints on the LFV $\tau$ couplings with
the scalar $\phi$. It is straightforward to find
\begin{equation} \label{eq:taulim}
  \begin{split}
    & \left| S^{e \tau} \right| < 5.9 \times 10^{-7} \, , \\
    & \left| S^{\mu \tau} \right| < 7.6 \times 10^{-7} \, .
  \end{split}
\end{equation}

These limits for the LFV couplings to $\tau$ leptons are expected to
be improved at Belle II. In fact, new methods for $\tau \to \ell \,
\phi$ searches at this experiment have been recently
proposed~\cite{DeLaCruz-Burelo:2020ozf}.

\subsection[\texorpdfstring{$\ell_\alpha \to \ell_\beta \, \gamma \, \phi$}{\unichar{"2113}\unichar{"03B1} \unichar{"2192} \unichar{"2113}\unichar{"03B2} \unichar{"03B3} \unichar{"03D5}} at the MEG experiment]{$\boldsymbol{\ell_\alpha \to \ell_\beta \, \gamma \, \phi}$ at the MEG experiment}

\begin{figure}[!tb]
  \centering
  \includegraphics[width=0.82\linewidth]{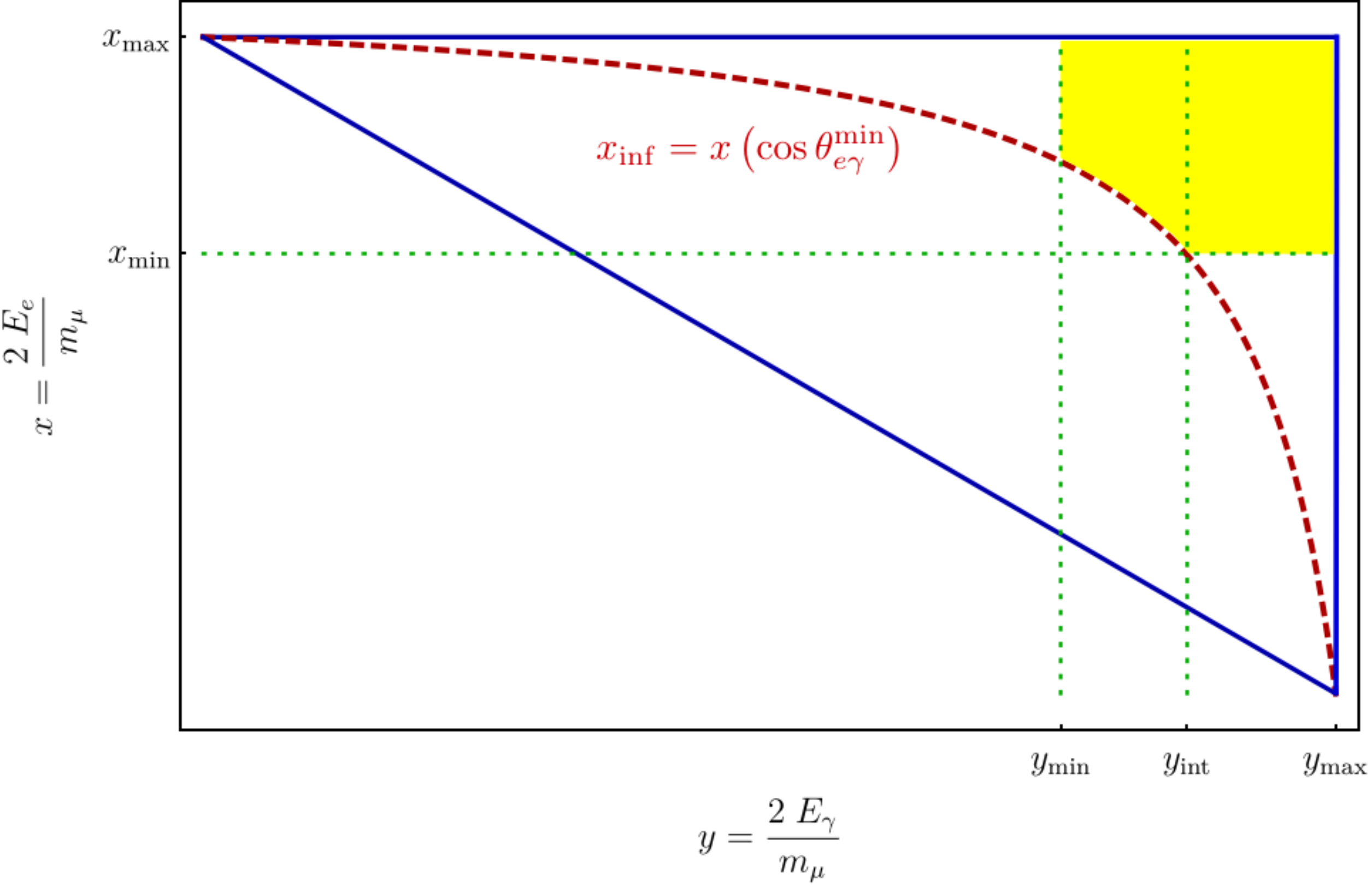}
  \caption{Illustration of the allowed phase space region for the
    process $\mu \to e \, \gamma \, \phi$ in a given experiment. The
    blue continuous lines correspond to $\cos \theta_{e \gamma} = \pm
    1$ and therefore delimit the total phase space that would be in
    principle available due to kinematics. The red dashed line
    represents $x_{\rm inf}(y)$ and corresponds to the minimal
    $\theta_{e \gamma}$ angle measurable by the experiment, excluding
    the region below it. The green dotted straight lines at $x_{\min}$
    and $y_{\min}$ are the minimal positron and photon energy,
    respectively, that the experiment can measure, while $y_{\rm int}$
    is the value of $y$ for which $x_{\min}$ and $x_{\rm inf}$
    intersect. Finally, the yellow surface is the region where we must
    integrate.
    \label{fig:phase-space}
    }
\end{figure}

\begin{figure}[tb]
  \begin{subfigure}{0.52\textwidth}
    \centering
    \includegraphics[width=1\linewidth]{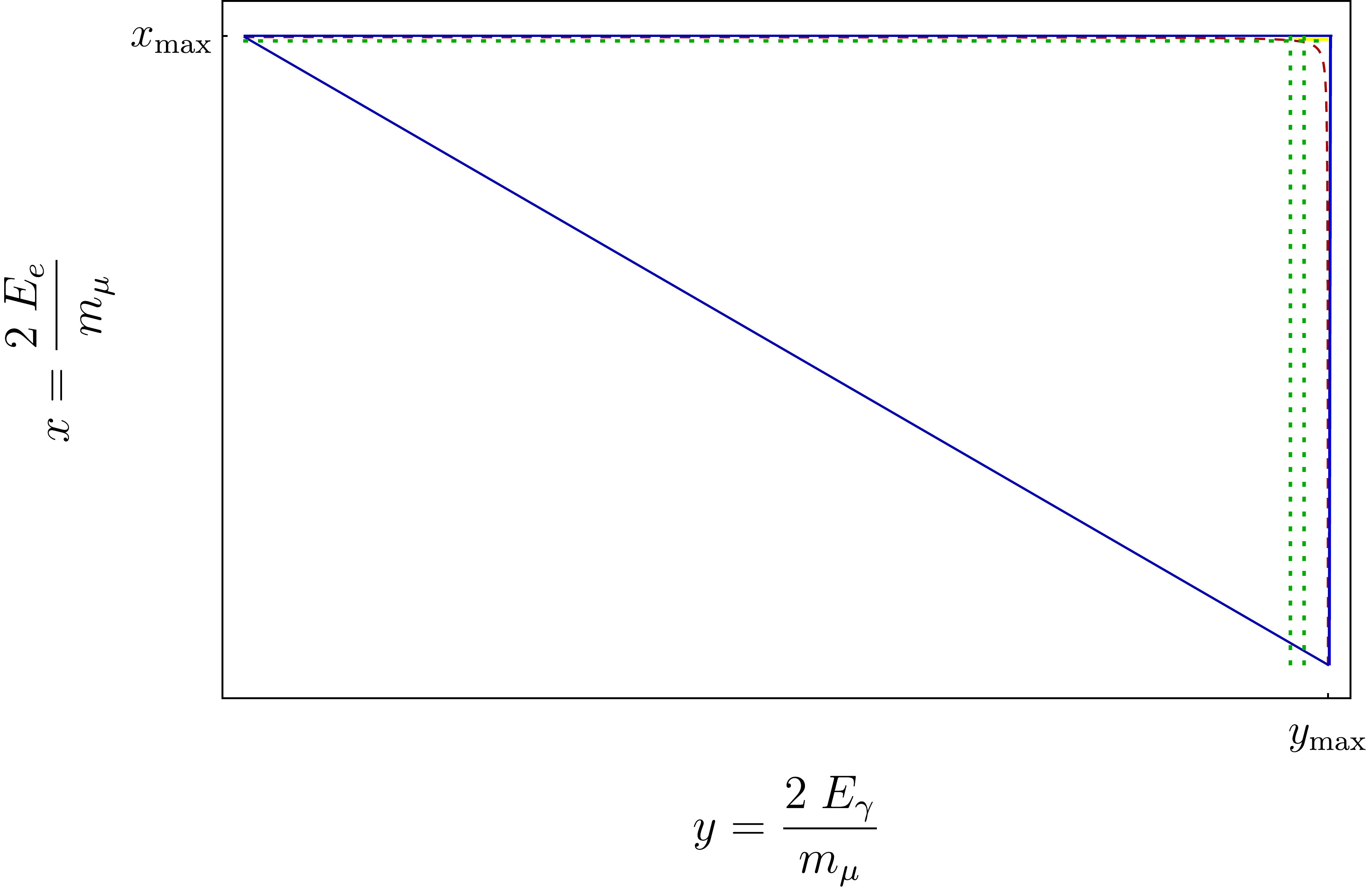}
  \end{subfigure}
  \begin{subfigure}{0.48\textwidth}
    \centering
    \includegraphics[width=1\linewidth]{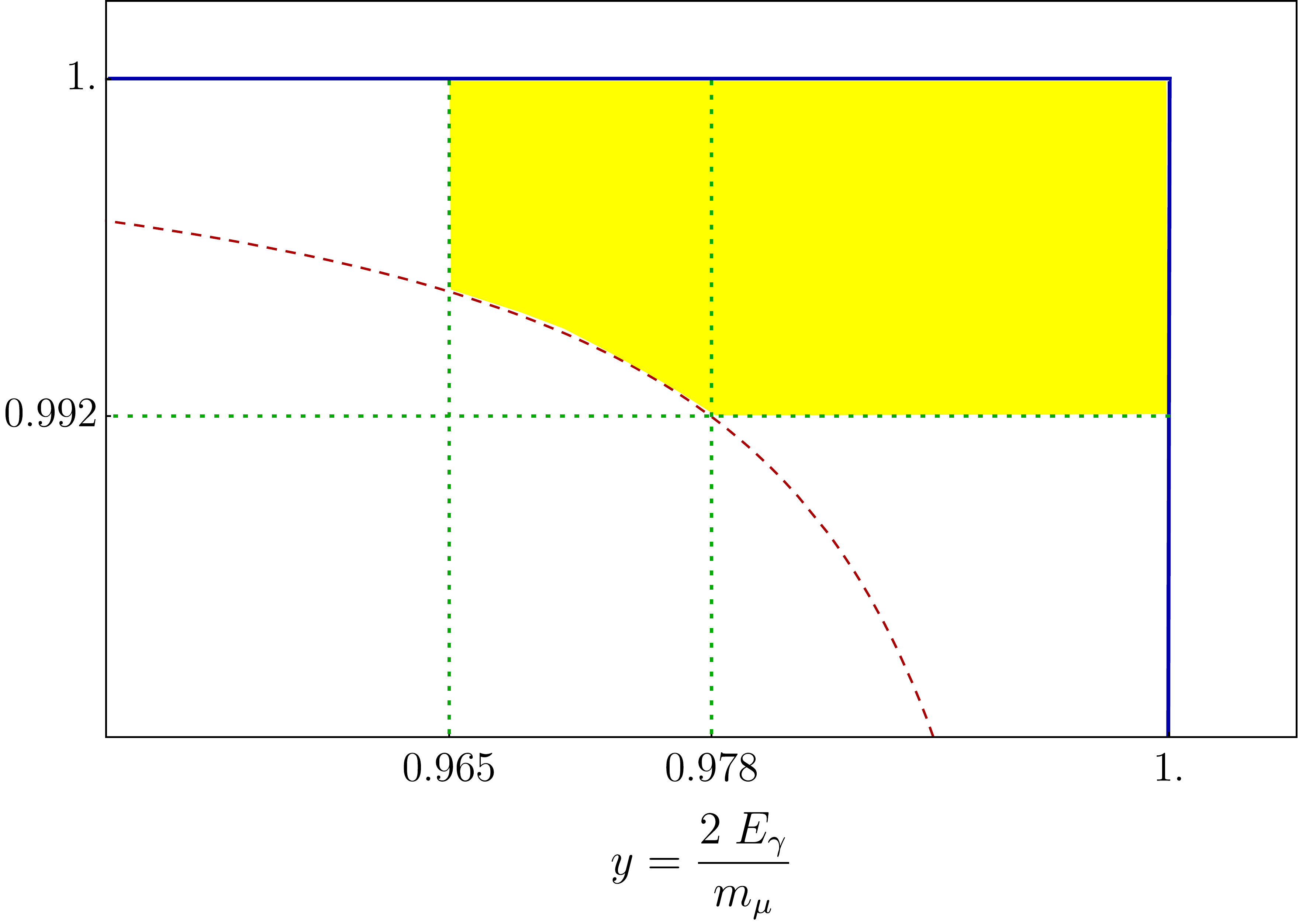}
  \end{subfigure}
  \caption{Realistic version of the phase space region limited by the
    experimental cuts of the MEG experiment, given in
    Eq.~\eqref{eq:MEGcuts}. The figure on the right shows a zoom of
    the figure on the left, centered on the colored surface.}
  \label{fig:phase-space-realistic}
\end{figure}

In order to illustrate the calculation of the phase space integral for
a specific case, let us focus on the $\mu \to e \, \gamma \, \phi$
decay and consider the MEG experiment~\cite{Mori:2016vwi}. This
experiment has been designed to search for $\mu \to e \, \gamma$ and
therefore concentrates on $E_e \simeq m_\mu / 2$ and $\cos \theta_{e
  \gamma} \simeq -1$ (positron and photon emitted back to
back). However, due to the finite experimental resolution, these cuts
cannot be imposed with full precision, which makes MEG also sensitive
to $\mu \to e \, \gamma \, \phi$. The final MEG results were obtained
with the cuts~\cite{Mori:2016vwi}
\begin{equation} \label{eq:MEGcuts}
  \cos \theta_{e \gamma} < - 0.99963 \quad, \quad 51.0 < E_\gamma < 55.5 \, \text{MeV} \quad, \quad 52.4 < E_e < 55.0 \, \text{MeV} \, .
\end{equation}
This defines the MEG kinematical region for the calculation of the
phase space integral in Eq.~\eqref{eq:phase_space_integral} since $\mu
\to e \, \gamma \, \phi$ events that fall in this region can be
detected by the experiment. For instance, events with $\cos \theta_{e
  \gamma} < - 0.99963$, or equivalently $\theta_{e \gamma} > \theta_{e
  \gamma}^{\min} = 178.441^{\circ}$, were at the reach of MEG. The
kinematical region can be divided into two subregions:
\begin{equation}
  \begin{split}
    y_{\min} = \frac{2 \, E_\gamma^{\min}}{m_\mu} < y & < y_{\rm int} \, , \\
    x_{\rm inf} < x & < x_{\max} = 1 \, ,
  \end{split}
\end{equation}
and
\begin{equation}
  \begin{split}
    y_{\rm int} < y & < y_{\max} = 1 \, , \\
    x_{\min} = \frac{2 \, E_e^{\min}}{m_\mu} < x & < x_{\max} \, ,
  \end{split}
\end{equation}
where $x_{\rm inf} = x_{\rm inf}(y)$ is the value of $x$ such that
$\cos \theta_{e \gamma} = \cos \theta_{e \gamma}^{\min}$ for each
value of $y$. This can be easily found by solving
Eq.~\eqref{eq:theta}:
\begin{equation}
  x_{\rm inf} = \frac{2 \, (1-y)}{2 - y \left(1 - \cos \theta_{e \gamma}^{\min}\right)} \, .
\end{equation}
Finally $y_{\rm int}$ is the value of $y$ for which $x_{\min}$ and $x_{\rm
  inf}$ coincide. These two subregions are illustrated in
Fig.~\ref{fig:phase-space}, where the experimental restrictions have
been modified for the sake of clarity by enlarging the kinematical
region of interest. A realistic representation obtained with the MEG
cuts in Eq.~\eqref{eq:MEGcuts} is shown in
Fig.~\ref{fig:phase-space-realistic}. This clearly illustrates the
strong suppression due to the phase space integral.

Having explained how to compute the phase space integral and
illustrated the strong suppression it introduces, we can obtain
results for the MEG experiment. Using the cuts in
Eq.~\eqref{eq:MEGcuts}, the phase space integral in
Eq.~\eqref{eq:phase_space_integral} can be numerically computed to
find
\begin{equation}
  \mathcal{I} \left( x_{ \min } , y_{ \min } \right)_{\rm MEG} = 3.8 \times 10^{-8} \, .
  \label{eq:phase_space_integra_num}
\end{equation}
Combining this result with Eq.~\eqref{eq:decaywidth_betagammaphi}, we
obtain the branching ratio of $\mu \to e \, \gamma \, \phi$ restricted
to the MEG phase space, obtaining
\begin{equation}
  \text{BR}_{\rm MEG} \left( \mu \to e \, \gamma \, \phi \right) = 1.5 \times 10^{5}  \left( \left| S^{e \mu}_L \right|^2 + \left| S^{e \mu}_R \right|^2 \right) \, .
  \label{eq:num_BRegammamu}
\end{equation}
MEG results require $\BR \left( \mu \to e \, \gamma \right) < 4.2
\times 10^{-13}$~\cite{Mori:2016vwi}, a bound that must also be
satisfied by $\text{BR}_{\rm MEG} \left( \mu \to e \, \gamma \, \phi
\right)$. This leads to
\begin{equation} \label{eq:limemu2}
	\left| S^{e \mu} \right| < 1.6 \times 10^{-9} \, .
\end{equation}
This bound is notably worse than the one given in
Eq.~\eqref{eq:limemu1}, as expected due to the strong phase space
suppression at MEG, an experiment that is clearly not designed to
search for $\mu \to e \, \gamma \, \phi$.

{
\renewcommand{\arraystretch}{1.2}
\begin{table}
\centering
\begin{tabular}{| c c c c c c c |}
  \hline
References & $\theta_{e \gamma}^{\min}$ & $E_\gamma^{\min}$ [MeV] & $E_e^{\min}$ [MeV] & $\mathcal{I} \left( x_{ \min } , y_{ \min } \right)$ & BR bound & Limit on $\left| S^{e \mu} \right|$ \\
\hline
\hline    
\cite{Bolton:1986tv} & $160^{\circ}$ & 40 & 44 & $1.3 \times 10^{-3}$ & $4.9 \times 10^{-11}$ & $9.5 \times 10^{-11}$ \\
\cite{Goldman:1987hy,Bolton:1988af} & $140^{\circ}$ & 38 & 38 & $1.1 \times 10^{-2}$ & $1.1 \times 10^{-9}$ & $1.6 \times 10^{-10}$ \\
\hline
\hline
\end{tabular}
\caption{Results in the search for $\mu \to e \, \gamma \, \phi$ at the Crystal Box experiment.
  \label{tab:crystalbox}
  }
\end{table}
}

More stringent bounds were obtained at the Crystal Box experiment at
LAMPF~\cite{Bolton:1986tv,Goldman:1987hy,Bolton:1988af}. Several
searches were performed, with different experimental cuts and
branching ratio bounds. These result in different limits on the
$\left| S^{e \mu} \right|$ effective coupling, as shown in
Table~\ref{tab:crystalbox}. Adapting the limit from the $\mu \to e
\gamma$ search in~\cite{Bolton:1986tv} along the lines followed in the
previous discussion for MEG, we find
\begin{equation}
  \left| S^{e \mu} \right| < 9.5 \times 10^{-11} \,  .
\end{equation}
This bound is still not better than the one given in
Eq.~\eqref{eq:limemu1}, but it is in the same ballpark. A very similar
bound is obtained with the results of a later analysis, in this case
more specific to $\mu \to e \, \gamma \,
\phi$~\cite{Goldman:1987hy,Bolton:1988af}.

Finally, the Mu3e experiment is not well equipped to detect the photon
in $\mu \to e \, \gamma \, \phi$ and therefore cannot improve on these
limits. As explained in~\cite{Perrevoortthesis}, a future
\textit{Mu3e-Gamma experiment} including a photon conversion layer
could increase the sensitivity to $\mu \to e \, \gamma \, \phi$.

\subsection[\texorpdfstring{$\ell_\alpha \to \ell_\beta \, \gamma$}{\unichar{"2113}\unichar{"03B1} \unichar{"2192} \unichar{"2113}\unichar{"03B2} \unichar{"03B3}} vs \texorpdfstring{$\ell_\alpha \to \ell_\beta \ell_\beta \ell_\beta$}{\unichar{"2113}\unichar{"03B1} \unichar{"2192} \unichar{"2113}\unichar{"03B2} \unichar{"2113}\unichar{"03B2} \unichar{"2113}\unichar{"03B2}}]{$\boldsymbol{\ell_\alpha \to \ell_\beta \gamma}$ vs $\boldsymbol{\ell_\alpha \to \ell_\beta \ell_\beta \ell_\beta}$}

The LFV decays $\ell_\alpha \to \ell_\beta \, \gamma$ and $\ell_\alpha
\to \ell_\beta \ell_\beta \ell_\beta$ constitute complementary probes of the
underlying physics. While $\ell_\alpha \to \ell_\beta \, \gamma$ only
receives contributions from dipole operators, $\ell_\alpha \to
\ell_\beta \ell_\beta \ell_\beta$ is induced by dipole as well as
non-dipole operators. Their relative importance can be studied by means of the ratio
\begin{equation} \label{eq:Rmue}
  R_{\alpha \beta} = \frac{\BR(\ell_\alpha \to \ell_\beta \ell_\beta
    \ell_\beta)}{\BR(\ell_\alpha \to \ell_\beta \, \gamma)} \, .
\end{equation}
In models in which the $\ell_\alpha \to \ell_\beta
  \ell_\beta \ell_\beta$ amplitude is clearly dominated by dipole
  contributions, the two branching ratios are strongly correlated and
  one can make a definite prediction for $R_{\mu e}$. In fact, since
  $\ell_\alpha \to \ell_\beta \ell_\beta \ell_\beta$ involves an
  additional electromagnetic coupling constant, one expects $R_{\alpha
    \beta} \ll 1$. Departures from this prediction would clearly point
  towards a non-dipole dominant contribution. We now consider this
  issue in the presence of an ultralight scalar, which contributes at
  tree-level to $\ell_\alpha \to \ell_\beta \ell_\beta \ell_\beta$ via
  scalar (and hence non-dipole) operators. Contrary to the
  above-mentioned dipole-dominated scenarios, in this case one
  generally expects $R_{\alpha \beta} \gg 1$, as shown below.

However, before we move on to the discussion of the interplay between
$\ell_\alpha \to \ell_\beta \, \gamma$ and $\ell_\alpha \to \ell_\beta
\ell_\beta \ell_\beta$, we would like to point out that light scalars
may offer additional experimental handles in $\ell_\alpha \to
\ell_\beta \ell_\beta \ell_\beta$. In particular, the authors
of~\cite{Heeck:2017xmg} showed that a light scalar produced on-shell
in $\ell_{\alpha}^{-} \to \ell_{\beta}^{-} \phi$ that later decays as
$\phi \to \ell_{\beta}^- \ell_{\beta}^+$ may lead to observable
displaced vertices. This interesting possibility is, however, not
possible in the ultralight scalar scenario considered here.

\subsubsection*{General dipole contributions}

First, we consider the general case of a
  scenario in which dipole contributions are independent of the
  non-dipole ones induced by the ultralight scalar $\phi$. This would
  be the case of a model containing additional LFV sources, not
  related to $\phi$. In order to evaluate the relevance of the new
contributions to $\ell_\alpha \to \ell_\beta \ell_\beta \ell_\beta$
mediated by the scalar $\phi$ we drop the 4-fermion operators in
Eq.~\eqref{eq:lag4F} and consider a simplified effective Lagrangian
containing only left-handed photonic dipole and scalar-mediated
operators
\begin{equation} \label{eq:lagSimp}
  \mathcal{L}_{\rm LFV}^{\rm simp} = \frac{e \, m_\alpha \, \left(K_2^L\right)^{\beta \alpha}}{2} \, \overline{\ell}_{\beta} \, \sigma^{\mu \nu} \, P_L \, \ell_{\alpha} F_{\mu \nu} + S_L^{\beta \alpha} \, \phi \, \overline{\ell}_\beta \, P_L \, \ell_\alpha + \hc \, .
\end{equation}
Then, inspired by~\cite{deGouvea:2013zba}, we parametrize the $K_2^L$
and $S_L$ coefficients as
\begin{equation} \label{eq:param}
  e \, \left(K_2^L\right)^{\beta \alpha} \equiv \frac{1}{\left(\kappa + 1 \right) \Lambda^2} \, , \qquad  S_L^{\beta \alpha} \equiv m_\alpha \, \frac{\kappa}{\left(\kappa + 1 \right) \Lambda} \, .
\end{equation}
$\Lambda$ is a dimensionful parameter that represents the energy scale
at which these coefficients are induced, while $\kappa$ is a
dimensionless parameter that accounts for the relative intensity of
these two interactions.~\footnote{We normalize $S_L$ by introducing
  the mass of the heaviest charged lepton involved in each
  process. However, this is done only for the purpose of this
  analysis. In the rest of the paper we do not assume any hierarchy
  among the couplings proportional to the charged lepton masses.} In
case of $\kappa \ll 1$, the dipole operator dominates, while the
scalar mediated contribution dominates for $\kappa \gg 1$. We point
out that $m_\alpha$ in Eqs.~\eqref{eq:lagSimp} and \eqref{eq:param} is
a global factor given by the mass of the heaviest charged lepton in
the process and that Eq.~\eqref{eq:param} assumes $S_L^{\beta \alpha}
= S_L^{\beta \beta}$.

\begin{figure}[tb]
  \centering
  \includegraphics[width=0.9\linewidth]{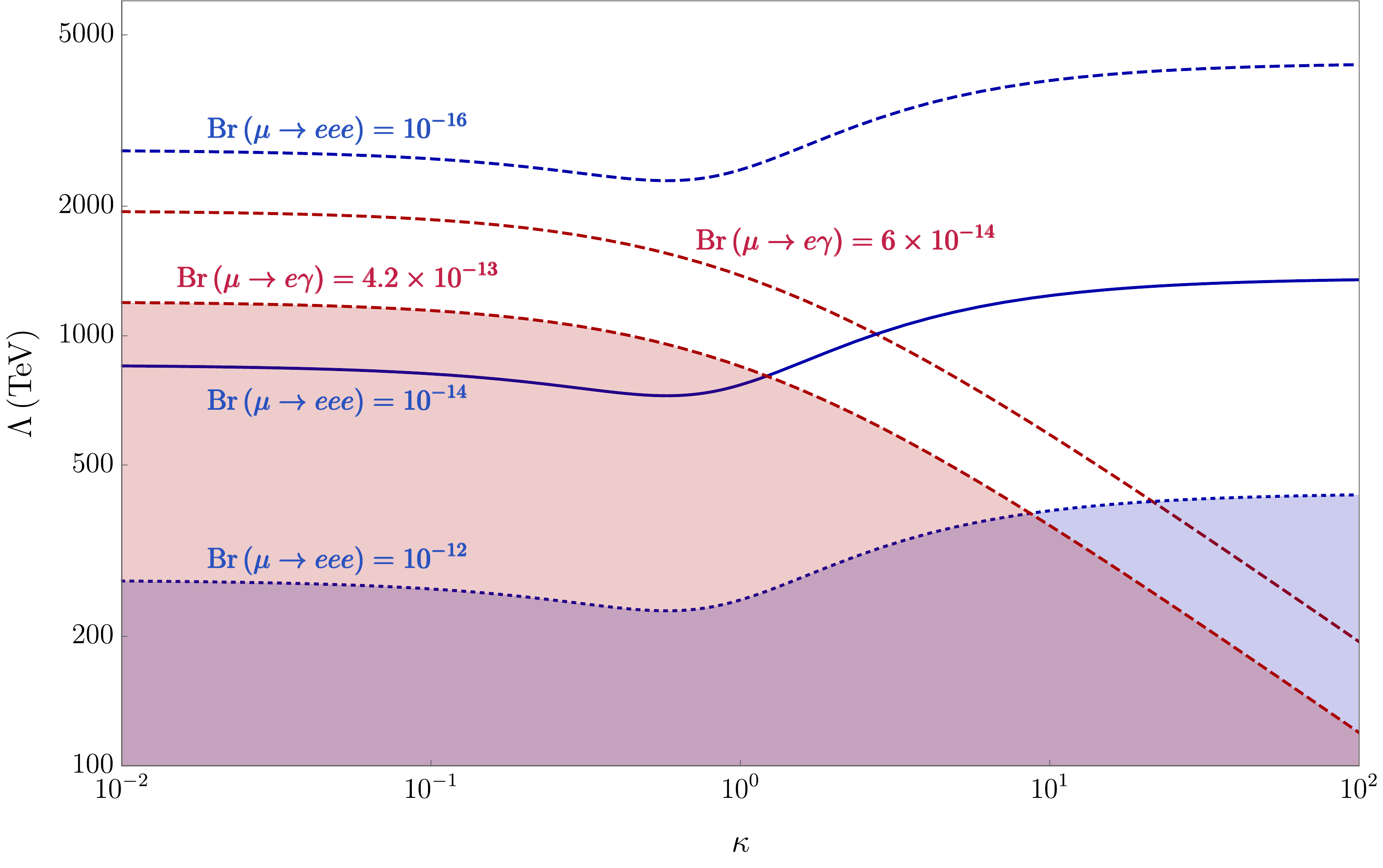}
  \caption{Contours of $\BR(\mu \to e \gamma)$ and $\BR(\mu \to eee)$
    in the $\kappa$-$\Lambda$ plane. The lowest values correspond to
    the future sensitivities for the MEG-II and Mu3e experiments,
    while colored regions are excluded due to the current bounds
    $\BR(\mu \to e \gamma) < 4.2 \cdot 10^{-13}$ and $\BR(\mu \to eee)
    < 10^{-12}$~\cite{Tanabashi:2018oca}. These results have been
    obtained with the effective Lagrangian in Eq.~\eqref{eq:lagS} and
    the parametrization in Eq.~\eqref{eq:param}.
    \label{fig:L3L}
  }
\end{figure}

\begin{figure}[tb]
  \begin{subfigure}{0.5\textwidth}
    \centering
    \includegraphics[width=1\linewidth]{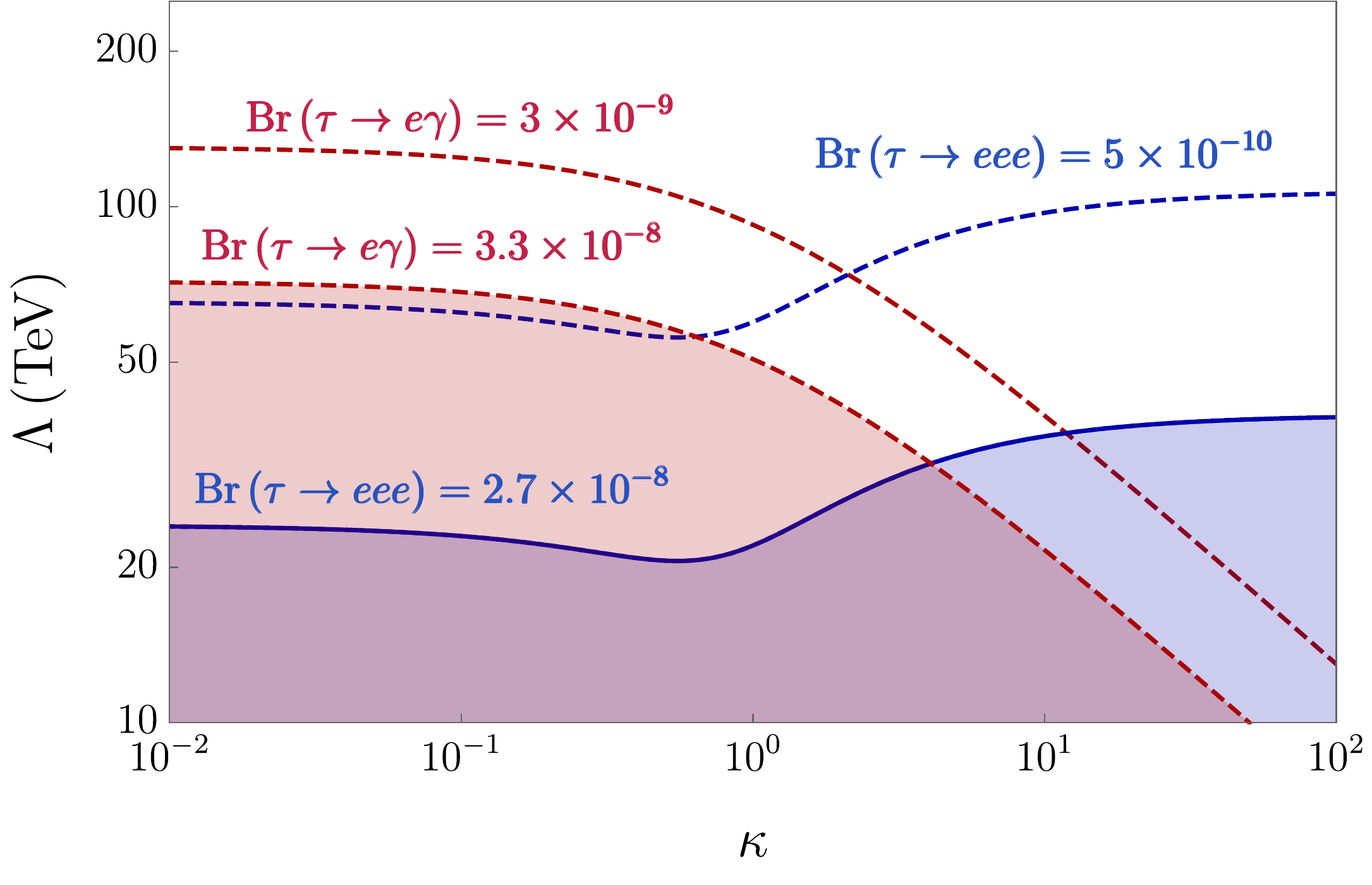}
  \end{subfigure}
  \begin{subfigure}{0.5\textwidth}
    \centering
    \includegraphics[width=1\linewidth]{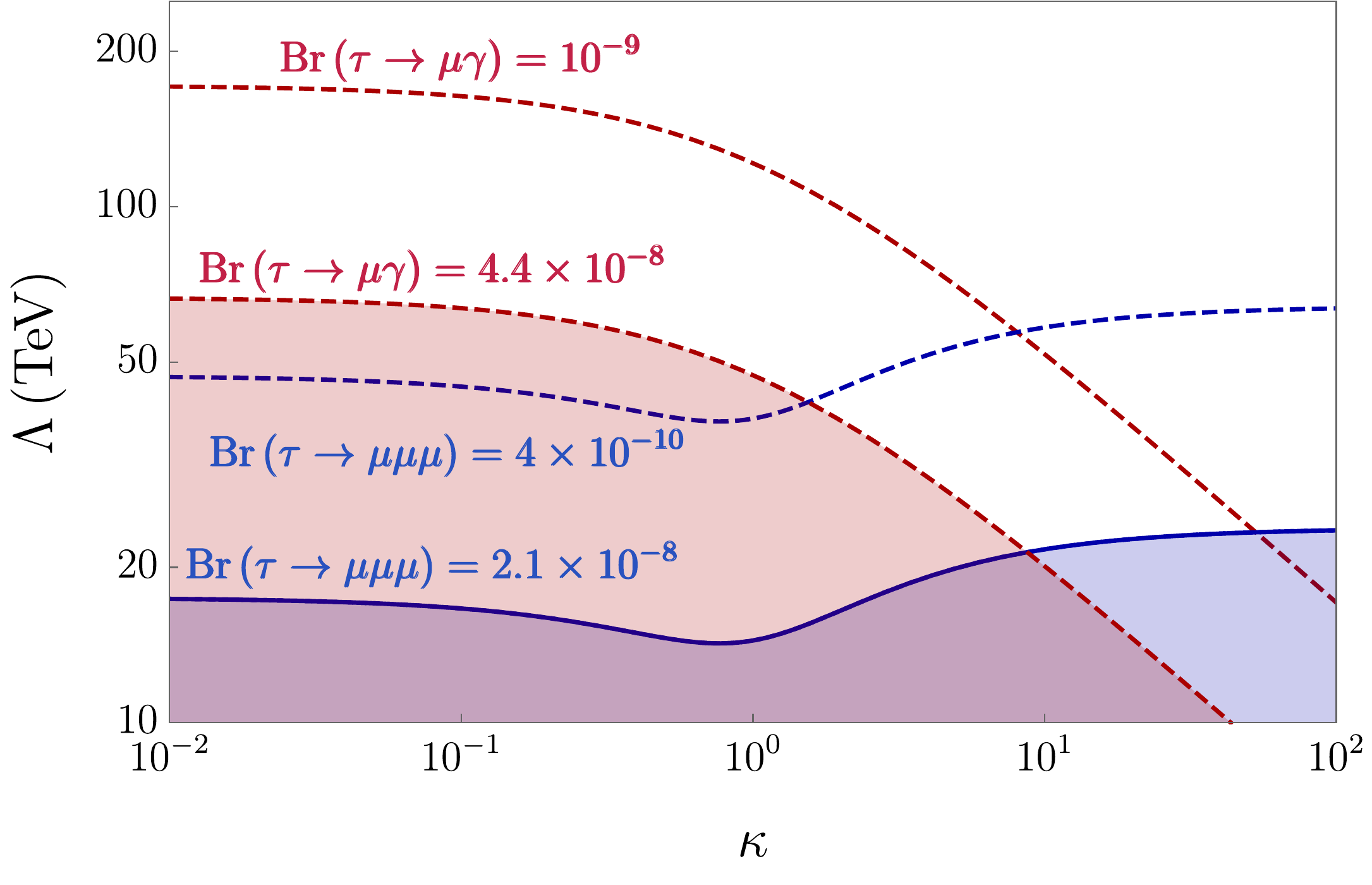}
  \end{subfigure}
  \caption{Contours of $\BR(\tau \to e \gamma)$ and $\BR(\tau \to
    eee)$, on the left, and $\BR(\tau \to \mu \gamma)$ and $\BR(\tau
    \to \mu \mu \mu)$, on the right, in the $\kappa$-$\Lambda$
    plane. The lowest values correspond to the expected future
    sensitivities of the Belle II experiment~\cite{Perez:2019cdy},
    while colored regions are excluded due to the current bounds
    $\BR(\tau \to e \gamma) < 3.3 \cdot 10^{-8}$, $\BR(\tau \to \mu
    \gamma) < 4.4 \cdot 10^{-8}$, $\BR(\tau \to eee) < 2.7 \cdot
    10^{-8}$ and $\BR(\tau \to \mu \mu \mu) < 2.1 \cdot
    10^{-8}$~\cite{Tanabashi:2018oca}. These results have been
    obtained with the effective Lagrangian in Eq.~\eqref{eq:lagS} and
    the parametrization in Eq.~\eqref{eq:param}.
  \label{fig:L3L-tau}}
\end{figure}

Fig.~\ref{fig:L3L} shows $\BR(\mu \to e \gamma)$ and $\BR(\mu \to
eee)$ as a function of $\Lambda$ and $\kappa$. Our results are
compared to the current bounds and the future sensitivities for the
MEG-II and Mu3e experiments. We observe that for $\kappa \gg 1$ and
$\BR(\mu \to eee) > 10^{-16}$, $\Lambda$ must be necessarily below
$\sim 3000$ TeV. A slightly lower upper limit for $\Lambda$ is found
when $\kappa \ll 1$ and $\BR(\mu \to e \gamma) > 10^{-14}$. These are
precisely the final expected sensitivities in MEG-II and
Mu3e. Furthermore, we note that the search for the scalar mediated
contribution in Mu3e will actually be very constraining in all the
parameter space. Similar results are shown for $\tau$
  decays in Fig.~\ref{fig:L3L-tau}. In this case, the current
  experimental limits are expected to be improved by about one order
  of magnitude by the LHCb and Belle II collaborations, which will
  search for the $\tau \to \ell_\beta \gamma$ and $\tau \to \ell_\beta
  \ell_\beta \ell_\beta$ decays, with $\ell_\beta = e, \mu$. This
  figure has been obtained using the expected sensitivities by the
  Belle II experiment presented in~\cite{Perez:2019cdy}. We find that
  for low values of $\kappa$, i.e. $\kappa \ll 1$, the current limit
  on $\BR(\tau \to e \gamma)$ implies the non-observation of $\tau \to
  eee$ at Belle II. This would therefore require a larger value of
  $\kappa$, to enhance the relative weight of the 3-body
  decay. Qualitatively similar results are obtained for $\tau \to \mu$
  transitions.

\subsubsection*{$\boldsymbol{\phi}$-induced dipole contributions}

We now consider the generation of dipole operators by loops involving
the ultralight scalar $\phi$, as discussed in
Sec.~\ref{sec:obsbetagamma} and shown in
Fig.~\ref{fig:Diagrambeta_gamma}. In this scenario, we assume that
$\phi$ provides the dominant (or, of course, only) contribution to
dipole operators. For the sake of simplicity, the couplings $S^{ee}$
and $S_{L,R}^{e \mu}$ will be the only ones allowed to be different
from zero in the analysis that follows. They will also be taken to be
real. In this case, the general expressions for $K_2^L$ and $K_2^R$
given in Eqs.~\eqref{eq:K2L} and \eqref{eq:K2R} lead to
\begin{align}
  \left(K_2^L\right)^{e \mu} = &  \frac{S^{ee}}{96 \pi^2 \, m_\mu^3} \biggl\{ 3 \, m_\mu \, S_R^{e \mu} + m_e \left( -6 \, S_L^{e \mu} + 2 \, \pi^2 \, S_L^{e \mu} + 3 \, S_R^{e \mu} \right) \biggr. \nonumber \\
  &  \biggl. + \, 3 \, m_e S_L^{e \mu} \log \left( - \frac{m_e^2}{m_\mu^2} \right) \left[ 1 + \log \left( - \frac{m_e^2}{m_\mu^2} \right) \right] \biggr\} \, , 
\end{align}
\begin{align}
  \left(K_2^R\right)^{e \mu} = & \frac{S^{ee}}{96 \pi^2 \, m_\mu^3} \biggl\{ 3 \, m_\mu \, S_L^{e \mu} + m_e \left( -6 \, S_R^{e \mu} + 2 \, \pi^2 \, S_R^{e \mu} + 3 \, S_L^{e \mu} \right) \biggr. \nonumber \\ 
  & \biggl. + \, 3 \, m_e S_R^{e \mu} \log \left( - \frac{m_e^2}{m_\mu^2} \right) \left[ 1 + \log \left( - \frac{m_e^2}{m_\mu^2} \right) \right] \biggr\} \, ,
\end{align}
where we have expanded at first order in
  $m_e$. These expressions allow us to compute the $R_{\mu e}$ ratio
  defined in Eq.~\eqref{eq:Rmue}. Defining the mass ratio $r =
  \frac{m_\mu^2}{m_e^2}$, we do that for some simplified scenarios:
\begin{itemize}

\item Scenario 1: $S_L^{e \mu} = 0$ or $S_R^{e \mu} = 0$

\begin{equation}
  R_{\mu e}^{(1)} \approx \frac{4 \, \pi \, r}{3 \, \alpha} \, \frac{12 \, \log r - 53}{|\log (-r)|^4 + r} \approx 3.2 \cdot 10^4 \, .
\end{equation} 

\item Scenario 2: $S_L^{e \mu} = S_R^{e \mu}$

\begin{equation}
  R_{\mu e}^{(2)} \approx \frac{4 \, \pi \, r}{3 \, \alpha} \, \frac{12 \, \log r - 53}{|\log^2 (-r) + \sqrt{r}|} \approx 1.9 \cdot 10^4 \, .
\end{equation} 

\item Scenario 3: $S_L^{e \mu} = - S_R^{e \mu}$

\begin{equation}
  R_{\mu e}^{(3)} \approx \frac{4 \, \pi \, r}{3 \, \alpha} \, \frac{12 \, \log r - 53}{|\log^2 (-r) - \sqrt{r}|} \approx 1.1 \cdot 10^5 \, .
\end{equation} 

\end{itemize}

We find that $R_{\mu e} \gg 1$ in these
  scenarios. This, however, was expected, since $\ell_\alpha \to
  \ell_\beta \ell_\beta \ell_\beta$ is induced at tree-level by $\phi$
  exchange, while $\ell_\alpha \to \ell_\beta \, \gamma$ can only take
  place at loop order. More interestingly, different scenarios for the
  $\phi$ couplings lead to very different predictions for $R_{\mu
    e}$. This would in principle allow us to determine the nature of
  the scalar $\phi$ if positive signals are observed for both $\mu \to
  e \gamma$ and $\mu \to eee$, and both branching ratios can be
  experimentally determined.

\subsection{Lepton magnetic and electric dipole moments}

At present, there is a discrepancy between the experimental
determination of the electron and muon AMMs and their SM predicted
values~\cite{Aoyama:2012wj,Aoyama:2012wk,Laporta:2017okg,Aoyama:2017uqe,Bennett:2006fi,Jegerlehner:2009ry,Blum:2018mom}
\begin{align}
  \Delta a_e &= a_e^{\text{exp}} - a_e^{\text{SM}} = (-87 \pm 36) \times 10^{-14} \, , \\
  \Delta a_{\mu} &= a_\mu^{\text{exp}} - a_\mu^{\text{SM}} = (27.1 \pm 7.3) \times 10^{-10} \, ,
\end{align}
where
\begin{equation}
  a_\beta = \frac{g_\beta - 2}{2} \, .
\end{equation}
In the case of the muon anomalous magnetic moment, the deviation is at
the level of $\sim 4 \, \sigma$, whereas for the electron anomalous
magnetic moment the significance is a little lower, slightly below
$\sim 3 \, \sigma$.~\footnote{See also the very recent calculation of
  the hadronic vacuum polarization contribution by the
  Budapest-Marseilles-Wuppertal collaboration~\cite{Borsanyi:2020mff},
  which brings the SM prediction for the muon anomalous magnetic
  moment into agreement with the experimental measurement. However,
  this result seems to lead to tension with electroweak
  data~\cite{Crivellin:2020zul}.} While further measurements (and
possibly improved theoretical calculations) are required to fully
confirm these anomalies, these intriguing deviations can be
interpreted as a possible hint of new
physics~\cite{Lindner:2016bgg}. In particular, the sign difference
between $\Delta a_e$ and $\Delta a_\mu$ and the relatively large value
of $|\Delta a_e|$ may indicate the presence of new physics
contributions that do not scale with the square of the corresponding
charged lepton masses~\cite{Giudice:2012ms}. In what
  concerns the EDMs of the charged leptons, the SM predicts tiny
  values, well beyond the experimental prospects in the near
  future. Therefore, any measurement of a non-zero charged lepton EDM
  would be a clear indication of CP-violating new physics effects. The
  current best limits for the electron and muon EDMs
  are~\cite{Andreev:2018ayy,Bennett:2008dy}
\begin{align}
  |d_e| &< 1.1 \times 10^{-29} \, e \, \text{cm} \, , \\
  |d_\mu| &< 1.5 \times 10^{-19} \, e \, \text{cm} \, ,  
\end{align}
both at $95 \%$ C.L..

\begin{figure}[!tb]
  \centering
  \includegraphics[width=0.442\linewidth]{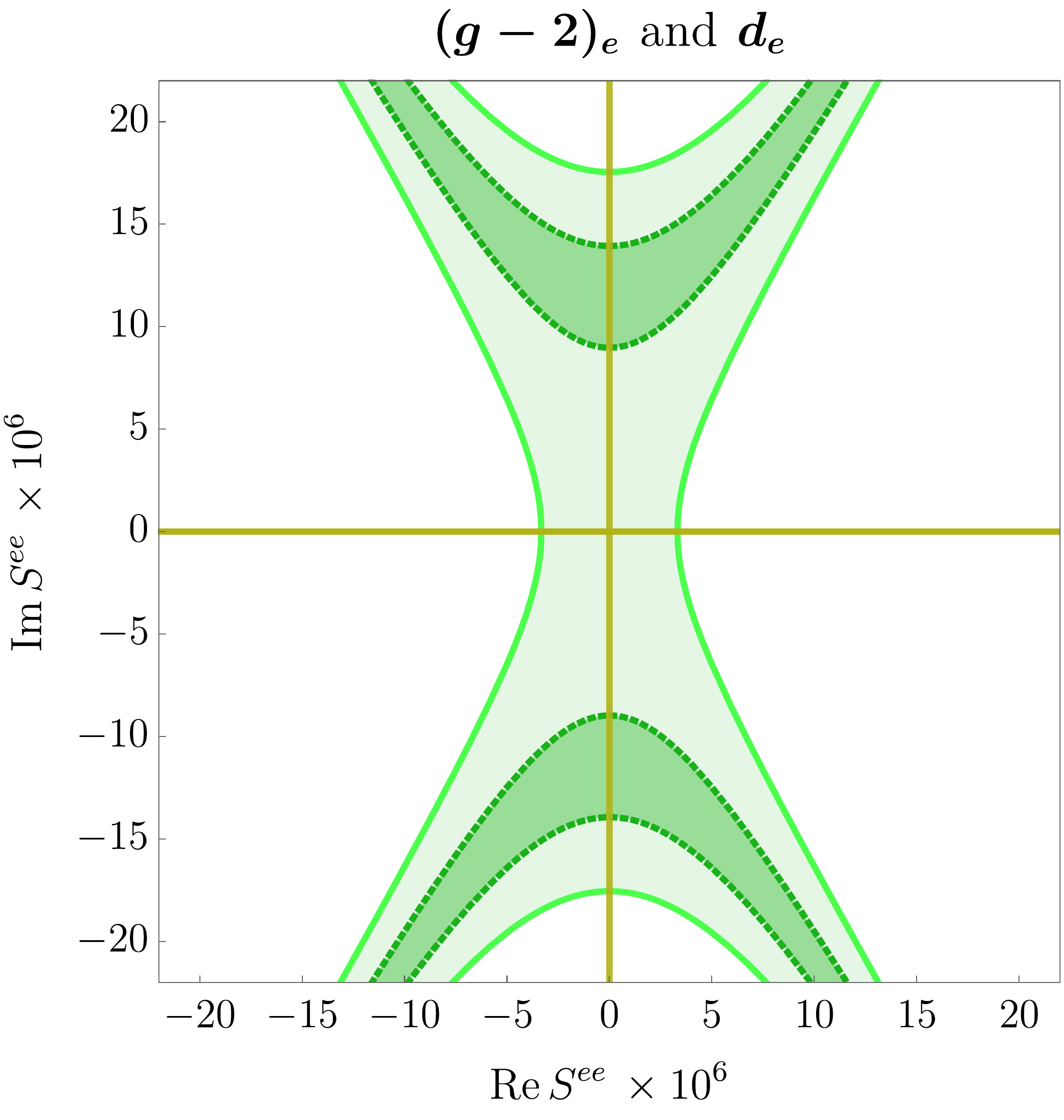}
  \hspace*{0.1cm}
  \includegraphics[width=0.45\linewidth]{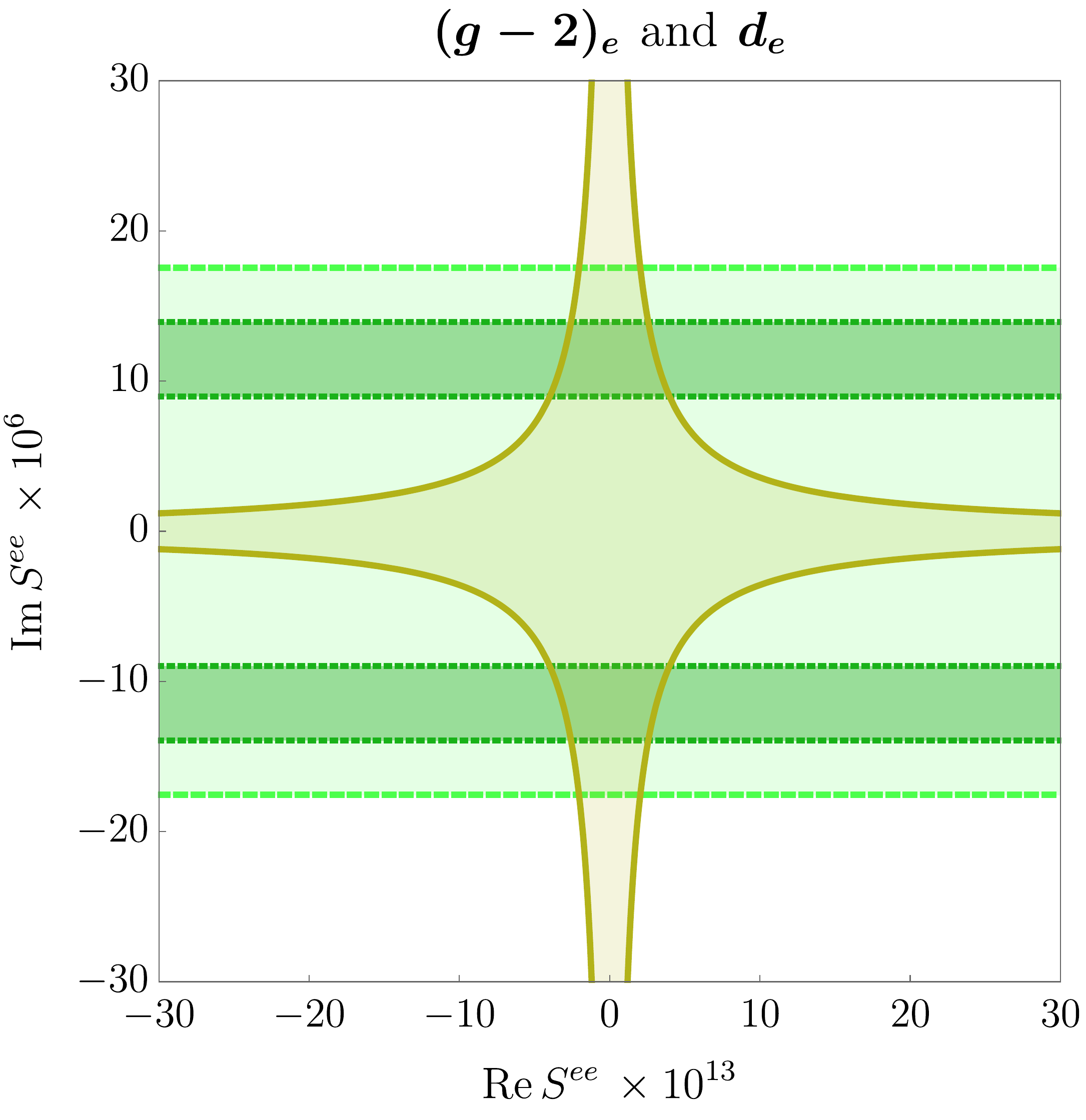}
  \caption{Favored region for the diagonal coupling $S^{e e}$, due to
    the electron anomalous magnetic and electric dipole moments.
    Within the light (dark) green region, the deviation in the
    electron AMM is explained at the $3\,\sigma$ ($1\,\sigma$)
    level. The region delimited by the orange continuous lines is the
    parameter space allowed by the current experimental upper bound of
    the electron EDM. In the figure on the right, the abscissa axis
    has been zoomed.
    \label{fig:eMoments}
    }
\end{figure}

\begin{figure}[!tb]
  \centering
  \includegraphics[width=0.45\linewidth]{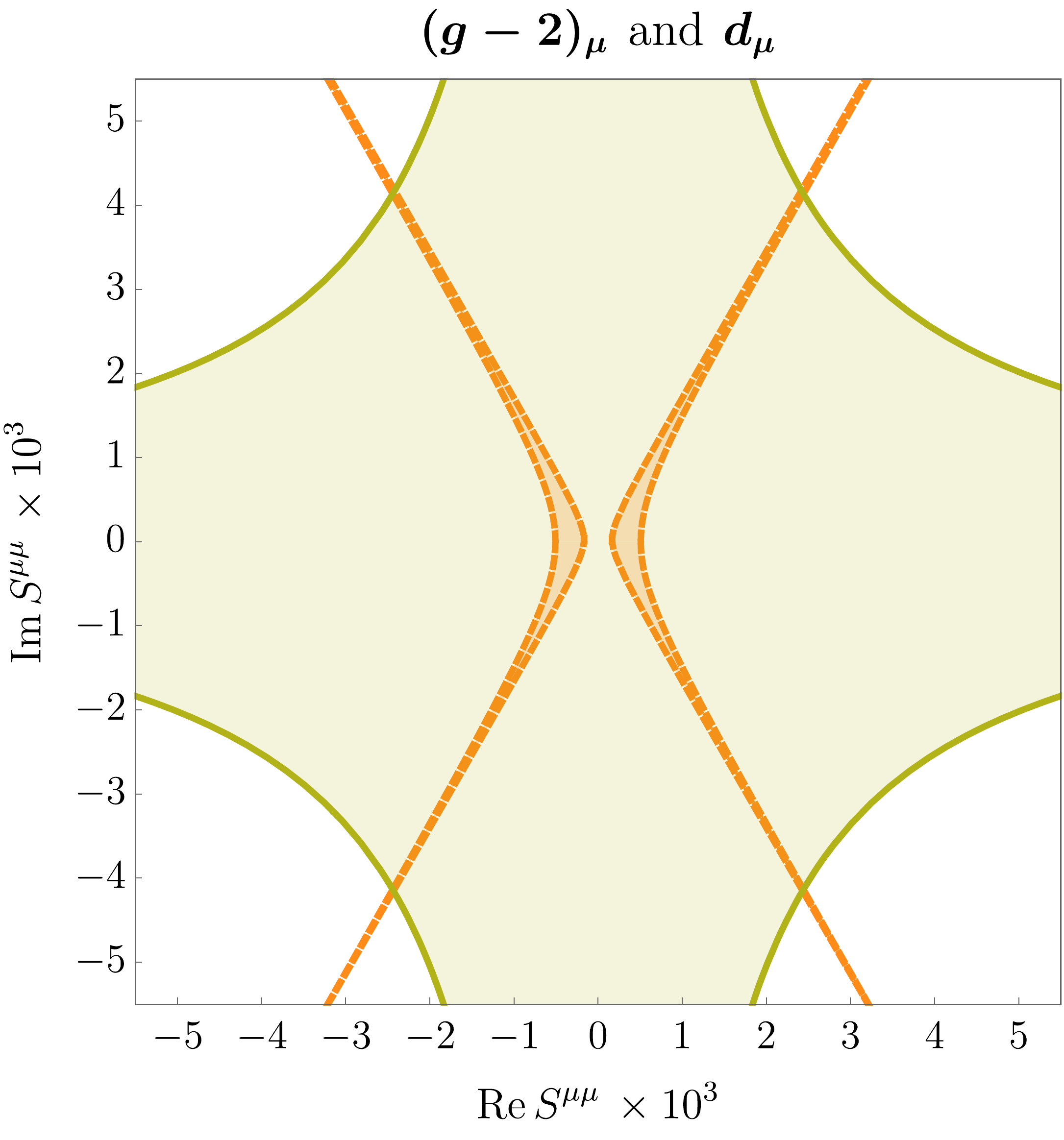}
  \hspace*{0.1cm}
  \includegraphics[width=0.45\linewidth]{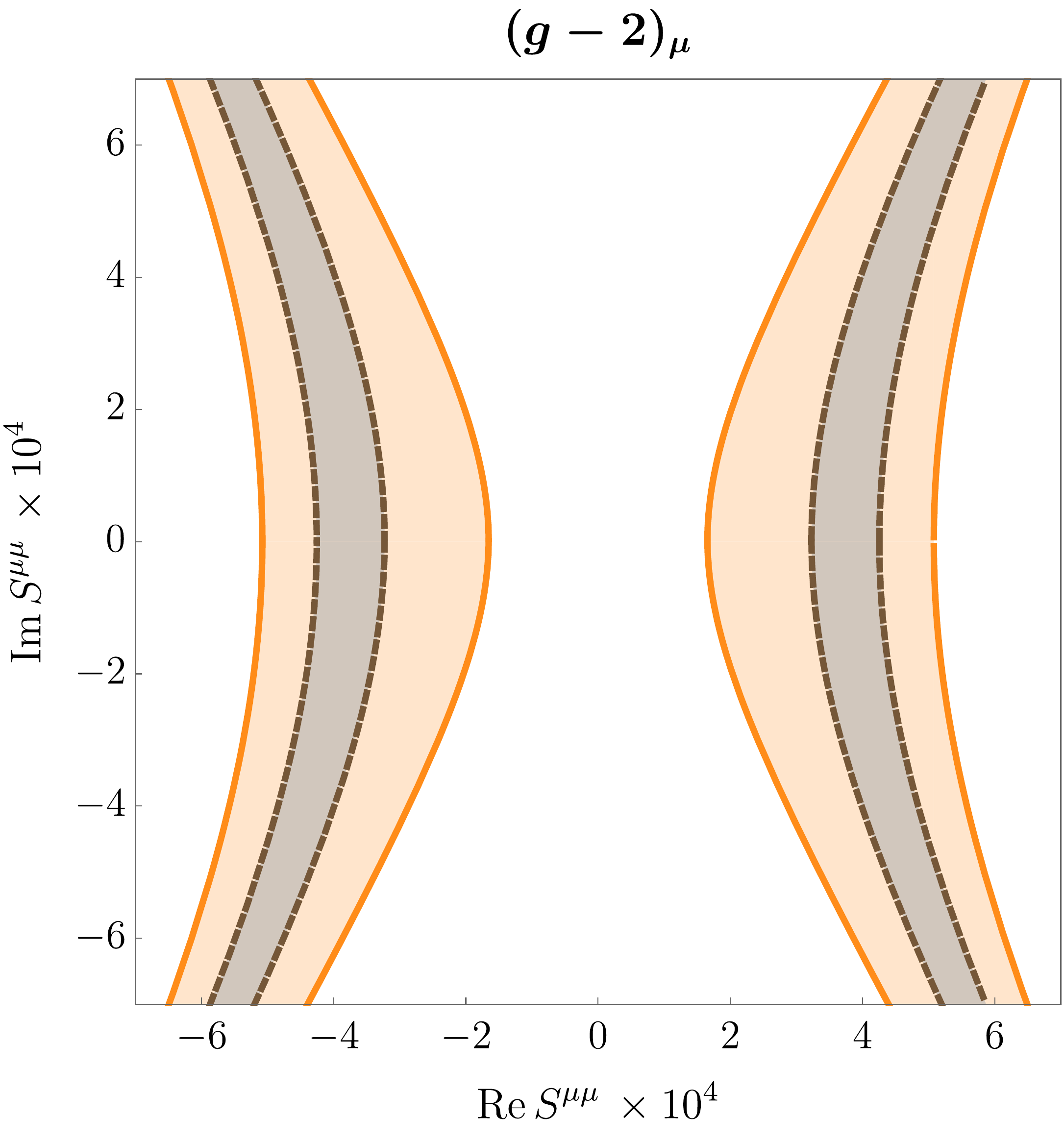}
  \caption{Favored regions for the diagonal coupling $S^{\mu\mu}$, due
    to the muon anomalous magnetic and electric dipole moments. In the
    figure on the left it is seen that the bound from the muon EDM
    (yellow continuous curves) does not restrict too much the AMM of
    the muon (orange dashed curves). On the right figure, only the
    muon AMM is represented and within the light (dark) region, the
    current experimental deviation is explained at the $3\,\sigma$
    ($1\,\sigma$) level.
    \label{fig:muMoments}
    }
\end{figure}

Figure~\ref{fig:eMoments} shows favored regions for
  the diagonal coupling $S^{ee}$ due to the electron AMM and EDM. As
  shown on the left panel, the bound on the electron EDM strongly
  constrains the $S^{ee}$ coupling, which must be essentially purely
  real or essentially purely imaginary. However, one can find regions
  in the parameter space that explain the $(g-2)_e$ anomaly,
  compatible with the bound on the electron EDM. Given the low
  significance of the $(g-2)_e$ anomaly, one stays within the $3 \,
  \sigma$ region even if $S^{ee} = 0$, but if $\text{Re} \, S^{ee}
  \lesssim 10^{-13}$, a value of about $\text{Im} \, S^{ee} \sim
  10^{-5}$ would actually achieve agreement at the $1\,\sigma$
  level. The deviation in $(g-2)_\mu$ is more significant, and this
  implies that one must introduce larger $S^{\mu\mu}$ values in order
  to reconcile the theoretical prediction with the experimental
  measurement. This is shown on Figure~\ref{fig:muMoments}. In this
  case, the bound from the muon EDM does not impose strong
  restrictions on the parameter space, as can be clearly seen in the
  left panel. However, larger $S^{\mu\mu}$ couplings, of the order of
  $10^{-4}$, are necessary in order to explain the current deviation
  between theory and experiment. In both cases, the required values
for $S^{ee}$ and $S^{\mu \mu}$ are in conflict with
the bounds discussed in Sec.~\ref{sec:bounds}, see Eqs.~\eqref{eq:See}
and \eqref{eq:Smumu}, and therefore a mechanism to suppress the
processes from which they are derived would be necessary for the
ultralight scalar $\phi$ to be able to provide an explanation to the
current $g-2$ anomalies.

Finally, we have explored whether the electron and
  muon AMM anomalies can be explained by purely off-diagonal
  contributions. In the following we consider vanishing diagonal
  couplings and real non-zero off-diagonal couplings. In this scenario
  the contribution to the charged leptons EDMs vanish and the AMMs
  strongly correlate with LFV observables. In fact, the bounds derived
  in Sec.~\ref{sec:AMM_EDM} from the non-observation of $\ell_\alpha
  \to \ell_\beta \, \phi$ imply that an explanation to the observed
  deviations cannot be achieved. In particular, we find that $S_L^{e
    \mu} \sim - S_R^{e \mu} \sim 2 \times 10^{-4}$ or $S_L^{e \tau}
  \sim - S_R^{e \tau} \sim 7 \times 10^{-4}$ are needed in order to
  explain the $\left( g - 2 \right)_e$ deviation. Regarding the muon
  AMM anomaly, only with the $\mu - \tau - \phi$ coupling one can
  obtain a positive contribution, requiring $S_L^{\mu \tau} \sim
  S_R^{\mu \tau} \sim 3 \times 10^{-3}$ to explain the $\left( g - 2
  \right)_\mu$ deviation. In all cases, the required off-diagonal
  couplings are several orders of magnitude larger than the limits in
  Eqs.~\eqref{eq:limemu1} and \eqref{eq:taulim}. We therefore conclude
  that the explanation of the electron and muon AMMs anomalies must
  come from diagonal contributions, whereas the off-diagonal ones can
  only play a subdominant role.

\section{Conclusions}
\label{sec:conclusions}

Ultralight scalars appear in a wide variety of SM extensions, either
as very light states or as exactly massless Goldstone bosons. Examples
include the axion and the majoron, two well-motivated hypothetical
particles at the core of two fundamental problems: the conservation of
CP in the strong interactions and the origin of neutrino masses. These
states, as well as other ultralight scalars, can be produced in many
leptonic processes or act as their mediators, leading to many exotic
signatures.

In this work we have explored the impact of ultralight scalars in many
leptonic observables. We have adopted a model independent general
approach, taking into account both scalar and pseudoscalar
interactions to charged leptons, therefore going beyond most existing
studies. First, we have briefly reviewed the current bounds from
stellar cooling, which set important constraints on the diagonal
couplings, and discussed indirect limits from the 1-loop generation of
a coupling to photons. Then, we have obtained
  analytical expressions for a wide variety of leptonic
  observables. We have revisited the decays $\ell_\alpha \to
\ell_\beta \, \phi$ and $\ell_\alpha \to \ell_\beta \, \gamma \,
\phi$, in which the scalar $\phi$ is produced, and provided complete
expressions for the radiative LFV decays $\ell_\alpha
  \to \ell_\beta \, \gamma$, as well as for the 3-body decays
$\ell_\alpha^- \to \ell_\beta^- \ell_\beta^- \ell_\beta^+$,
$\ell_\alpha^- \to \ell_\beta^- \ell_\gamma^- \ell_\gamma^+$ and
$\ell_\alpha^- \to \ell_\beta^+ \ell_\gamma^- \ell_\gamma^-$, in which
$\phi$ contributes as mediator. The effect of ultralight scalars on
the charged leptons anomalous magnetic and electric
  dipole moments has also be discussed. Finally,
  several phenomenological aspects of this scenario are
  explored. After deriving limits on off-diagonal couplings from
  lepton flavor violating observables, we have shown that an
  explanation to the $(g-2)_e$ and $(g-2)_\mu$ anomalies is possible
  in this scenario. We have also shown that the observables discussed
  in this paper are indeed complementary.

The phenomenology of ultralight scalars is very rich, since they are
kinematically accessible in most high- and low-energy processes. We
have discussed many purely leptonic processes, but if $\phi$ couples
to quarks as well, many hadronic and semi-leptonic channels open. This
could give rise to many signatures at kaon
factories~\cite{Gori:2020xvq}. Furthermore, ultralight scalars may
leave their footprints in other processes. For instance, they can be
produced and emitted in tritium beta decay~\cite{Arcadi:2018xdd} or
$\mu-e$ conversion in nuclei~\cite{Uesaka:2020okd}, have a strong
impact in leptogenesis~\cite{Sierra:2014sta}, and give rise to
non-resonant phenomena at colliders~\cite{Folgado:2020utn}. In our
opinion, this diversity of experimental signatures and their potential
to unravel some of the most important problems in particle physics
through their connection to ultralight scalars merits further
investigation.

\section*{Acknowledgements}

The authors are grateful to Julian Heeck, Mario Reig and Martin Hirsch
for fruitful discussions. Work supported by the Spanish grants
FPA2017-85216-P (MINECO/AEI/FEDER, UE), SEJI/2018/033 (Generalitat
Valenciana) and FPA2017-90566-REDC (Red Consolider MultiDark).  The
work of PE is supported by the FPI grant PRE2018-084599. AV
acknowledges financial support from MINECO through the Ramón y Cajal
contract RYC2018-025795-I.

\appendix

\section{Parametrization in terms of derivative interactions}
\label{sec:polar}

Eq.~\eqref{eq:lagS} is completely general and includes both scalar and
pseudoscalar interactions of the field $\phi$ with a pair of charged
leptons. An alternative parametrization in terms of derivative
interactions is given by
\begin{equation}
\mathcal{L}_{\ell \ell \phi} = \left( \partial_\mu \phi \right) \, \bar{\ell}_\beta \gamma^\mu \left( \widetilde{S}_L^{\beta \alpha} P_L + \widetilde{S}_R^{\beta \alpha} P_R \right) \ell_\alpha + \hc \, .
\label{eq:lagS2}
\end{equation}
The coefficients $\widetilde{S}_{L,R}$ have dimensions of mass$^{-1}$
and we consider all possible flavor combinations: $\beta \alpha =
\left\{ ee, \mu\mu,\tau\tau,e\mu,e\tau, \mu\tau\right\}$. Notice that
the diagonal $\ell_\beta - \ell_\beta - \phi$ vertex is proportional
to $(\widetilde{S}_L + \widetilde{S}_L^\ast)^{\beta \beta} P_L +
(\widetilde{S}_R + \widetilde{S}_R^\ast)^{\beta \beta} P_R$, and
therefore the diagonal couplings can be taken to be real without loss
of generality. As will be shown below, Eq.~\eqref{eq:lagS2} only
includes pseudoscalar interactions for $\phi$. Therefore, it can be
thought of as a particularization of
Eq.~\eqref{eq:lagS}.~\footnote{The parametrization in
  Eq.~\eqref{eq:lagS2} is completely general if $\phi$ is a pure
  pseudoscalar, usually the case of the Goldstone bosons in many
  models. In such scenarios, the two parametrizations for the
  effective Lagrangian $\mathcal{L}_{\ell \ell \phi}$ introduced here
  are related to two possible ways to parametrize the Goldstone
  boson. Eq.~\eqref{eq:lagS} follows from a \textit{cartesian
    parametrization}, that splits a complex scalar field in terms of
  its real and imaginary components. Alternatively, the
  parametrization in terms of derivative interactions in
  Eq.~\eqref{eq:lagS2} would follow from a \textit{polar
    parametrization}, that splits a complex scalar field in terms of
  its modulus and phase. As we will prove below, they lead to the same
  results for observables involving on-shell leptons.}

Physical observables must be independent of the parametrization
chosen. We proceed to show now that the two parametrizations
considered here are completely equivalent for a pure
  pseudoscalar in processes involving on-shell leptons. First, we
recall the equations of motion for the lepton fields $\ell_\alpha$ and
its conjugate $\bar{\ell}_\alpha$
\begin{equation}
  \begin{split}
    & i \, \gamma^\mu \partial_\mu \ell_\alpha - m_\alpha \ell_\alpha = 0 \, ,  \\
    & i \, \partial_\mu \bar{\ell}_\alpha \gamma^\mu + m_\alpha \bar{\ell}_\alpha = 0 \, ,
  \end{split}
  \label{eq:eom}
\end{equation}
valid for on-shell leptons. One can now rewrite
Eq.~\eqref{eq:lagS2} as the sum of a total derivative and a derivative
acting on the lepton fields. The total derivative does not contribute
to the action, whereas the derivative on the lepton fields can be
replaced using the equations of motion in Eq.~\eqref{eq:eom}. This
leads to
\begin{equation}
  \begin{split}
    \mathcal{L}_{\ell \ell \phi} & = - i \, \phi \, \bar{\ell}_\beta \left[ \left( m_\beta \, \widetilde{S}^{\beta \alpha}_L - m_\alpha \, \widetilde{S}^{\beta \alpha}_R \right) P_L + \left( m_\beta \, \widetilde{S}^{\beta \alpha}_R - m_\alpha \, \widetilde{S}^{\beta \alpha}_L \right) P_R \right] \ell_\alpha + \hc \\
    & \equiv \phi \, \bar{\ell}_\beta \left( S^{\beta \alpha}_L P_L + S^{\beta \alpha}_R P_R \right) \ell_\alpha + \hc \, . \\
  \end{split}
  \label{eq:equiv}
\end{equation}
Therefore we find a \textit{dictionary} between the $S_X$ and
$\widetilde{S}_X$ coefficients
\begin{align}
  S_L^{\beta \alpha} & = i \left( m_\alpha \, \widetilde{S}_R^{\beta \alpha} - m_\beta \, \widetilde{S}_L^{\beta \alpha} \right) \, , \label{eq:dic1} \\
  S_R^{\beta \alpha} & = i \left( m_\alpha \, \widetilde{S}_L^{\beta \alpha} - m_\beta \, \widetilde{S}_R^{\beta \alpha} \right) \, , \label{eq:dic2}
\end{align}
which for the diagonal couplings reduces to
\begin{align}
  S^{\beta \beta} & = S_L^{\beta \beta} + S_R^{\beta \beta \ast} = 2 \, i \, m_\beta \left( \widetilde{S}_R^{\beta \beta} - \widetilde{S}_L^{\beta \beta} \right) \, . \label{eq:dic3} 
\end{align}
Since both $\widetilde{S}_X^{\beta \beta}$ are real parameters,
Eq.~\eqref{eq:dic3} implies that the diagonal $S^{\beta \beta}$
couplings must be purely imaginary. It is straightforward to show
that, in this case, the flavor conserving interactions of $\phi$ in
Eq.~\eqref{eq:lagS} are proportional to $\gamma_5$ (see
Eq.~\eqref{eq:diag}). This proves that Eq.~\eqref{eq:lagS2} is not
general, but only includes pseudoscalar interactions, and there is no
one-to-one correspondence between the two parametrizations. Given a
set of $\widetilde{S}_X$ couplings, one can always find the
corresponding $S_X$ couplings using Eqs.~\eqref{eq:dic1} and
\eqref{eq:dic2}. However, certain sets of $S_X$ couplings, namely
those with non-vanishing real parts, cannot be expressed in terms of
$\widetilde{S}_X$ couplings. This stems from the fact that purely
scalar interactions are not included in Eq.~\eqref{eq:lagS2}.

The equivalence for the case of a pure pseudoscalar can be explicitly
illustrated by comparing the analytical expressions obtained with
Eqs.~\eqref{eq:lagS} and \eqref{eq:lagS2} for a given observable. We
can start with a trivial example, the process $\ell_\alpha \to
\ell_\beta \phi$, discussed in Sec.~\ref{sec:obsbetaphi}. Using the
parametrization in Eq.~\eqref{eq:lagS2}, one can easily derive the
decay width of this two-body decay,
\begin{equation}
  \widetilde{\Gamma} \left(\ell_\alpha \to \ell_\beta \, \phi \right) = \frac{m_\alpha^3}{32 \, \pi} \left( \left| \widetilde{S}_L^{\beta \alpha} \right|^2 + \left| \widetilde{S}_R^{\beta \alpha} \right|^2 \right) \, ,
  \label{eq:decaywidth_betaphi_der}
\end{equation}
where terms proportional to $m_\beta$ have been neglected. This
results differs from Eq.~\eqref{eq:decaywidth_betaphi} only by a factor
$m_\alpha^2$, as one would obtain from the direct application of the
dictionary in Eqs.~\eqref{eq:dic1} and \eqref{eq:dic2}. Let us now
consider a less trivial example: $\ell_\alpha^- \to \ell_\beta^-
\ell_\beta^+ \ell_\beta^-$. The computation of its amplitude with the
Lagrangian in Eq.~\eqref{eq:lagS2} makes use of the same Feynman
diagrams shown in Fig.~\ref{fig:Diagrams}. In this case one obtains
\begin{equation}
	\begin{split}
	\widetilde{\mathcal{M}}_\phi &= \bar{u} \left( p_3 \right) 2 \left( - \slashed{q} \right) \left( \widetilde{S}_L^{\beta \beta} P_L + \widetilde{S}_R^{\beta \beta} P_R \right) v \left(p_4\right) \frac{i}{q^2 + i \varepsilon} \bar{u} \left(p_2\right)  \left( \slashed{q} \right) \left( \widetilde{S}_L^{\beta \alpha} P_L + \widetilde{S}_R^{\beta \alpha} P_R \right) u \left( p_1 \right) \\
	& - \bar{u} \left( p_2 \right) 2 \left( - \slashed{k} \right) \left( \widetilde{S}_L^{\beta \beta} P_L + \widetilde{S}_R^{\beta \beta} P_R \right) v \left(p_4\right) \frac{i}{k^2 + i \varepsilon} \bar{u} \left(p_3\right)  \left( \slashed{k} \right) \left( \widetilde{S}_L^{\beta \alpha} P_L + \widetilde{S}_R^{\beta \alpha} P_R \right) u \left( p_1 \right) \, ,
\end{split}
\label{eq:AmplitudeS2}
\end{equation}
where the factor of $2$ preceding the diagonal coupling is due to the
addition of the Hermitian conjugate, as explicitly shown in
Eq.~\eqref{eq:lagS2}. Again, explicit flavor indices have been
introduced. The decay width is computed to be
\begin{equation}
  \begin{split}
    & \widetilde{\Gamma}_{\phi}\left( \ell_\alpha^- \to \ell_\beta^- \ell_\beta^+ \ell_\beta^- \right) = \\
    & \frac{m_{\alpha}^{5}}{512 \pi^{3}} \Biggl\{4 \left( \left|\widetilde{S}_{L}^{\beta \alpha}\right|^{2} + \left|\widetilde{S}_{R}^{\beta \alpha}\right|^{2} \right) \left( \widetilde{S}_L^{\beta \beta} - \widetilde{S}_R^{\beta \beta} \right)^2 \frac{m_\beta^2}{m_\alpha^2} \left( 4 \log \frac{m_\alpha}{m_\beta} - \frac{15}{2} \right) \Biggr. \\
    & + \frac{m_\beta}{3 \, m_\alpha} \biggl\{ \left( \widetilde{S}_L^{\beta \beta} - \widetilde{S}_R^{\beta \beta} \right) \left\{ \widetilde{S}^{\beta \alpha}_R \left( A^{S *}_{LL} - 2 A^{S *}_{LR} \right) - \widetilde{S}^{\beta \alpha}_L \left( A^{S *}_{RR} - 2 A^{S *}_{RL} \right) \right. \biggr. \\
    & + \frac{m_\beta}{m_\alpha} \left\{ \widetilde{S}_{L}^{\beta \alpha} \left[ 2 A^{S *}_{LL} + \left(12 \log \frac{m_\alpha}{m_\beta} - 25 \right) A^{S *}_{LR} \right] - \widetilde{S}_{R}^{\beta \alpha} \left[ 2 A^{S *}_{RR} + \left(12 \log \frac{m_\alpha}{m_\beta} - 25 \right) A^{S *}_{RL} \right] \right\} \\
    & + 12 \left[ A^{T *}_{RR} \left( \widetilde{S}^{\beta \alpha}_{L} + 2 \frac{m_\beta}{m_\alpha} \widetilde{S}^{\beta \alpha}_R \right) - A^{T *}_{LL} \left( \widetilde{S}^{\beta \alpha}_{R} + 2 \frac{m_\beta}{m_\alpha} \widetilde{S}^{\beta \alpha}_L \right) \right]  + 4 \left( \widetilde{S}^{\beta \alpha}_R A^{V *}_{LR} - \widetilde{S}^{\beta \alpha}_L A^{V *}_{RL}  \right) \\
    & + 2 \frac{m_\beta}{m_\alpha} \left\{ \widetilde{S}^{\beta \alpha}_L \left[ \left( 25 - 12 \log \frac{m_\alpha}{m_\beta} \right) A^{V *}_{LR} - \left( 42 - 24 \log \frac{m_\alpha}{m_\beta} \right) A^{V *}_{LL} \right] \right. \\
    & - \left. \widetilde{S}^{\beta \alpha}_R \left[ \left( 25 - 12 \log \frac{m_\alpha}{m_\beta} \right) A^{V *}_{RL} - \left( 42 - 24 \log \frac{m_\alpha}{m_\beta} \right) A^{V *}_{RR} \right] \right\} \\
    & + 6 e^2 \left[ \left(K^L_2\right)^{\beta \alpha \ast} \widetilde{S}^{\beta \alpha}_L - \left(K^R_2\right)^{\beta \alpha \ast} \widetilde{S}^{\beta \alpha}_R \right] \\
    & \Biggl. \biggl. \left. + \, 4 e^2 \frac{m_\beta}{m_\alpha} \left(\frac{3}{2} + \pi^2 + 6 \log^2 2 - 6 \log^2 \frac{m_\alpha}{m_\beta} \right) \left[ \left(K^R_2\right)^{\beta \alpha \ast} \widetilde{S}^{\beta \alpha}_L - \left(K^L_2\right)^{\beta \alpha \ast} \widetilde{S}^{\beta \alpha}_R \right] \right\} + \cc \biggr\} \Biggr\},
	\end{split}
	\label{eq:widthS2}
\end{equation}
where in this expression $A_{X Y}^I = \left( A_{X Y}^I \right)^{\beta
  \beta \beta \alpha}$. We note that infrarred divergences also occur
in interference terms at this order in
$\frac{m_\beta}{m_\alpha}$. This explains the appearance of several
log factors. The decay width in Eq.~\eqref{eq:widthS2} can be compared
to a previous result in the literature. The authors
of~\cite{Bjorkeroth:2018dzu} drop all interference terms in their
calculation, and then their result must be compared to the first line
in Eq.~\eqref{eq:widthS2}. One can easily relate the
$\widetilde{S}_{L,R}$ coefficients to the ones
in~\cite{Bjorkeroth:2018dzu} as
\begin{equation}
  V^e_{\beta \alpha} \equiv - \frac{1}{2} \left( \widetilde{S}^{\beta \alpha}_L + \widetilde{S}^{\beta \alpha}_R \right) \, , \qquad A^e_{\beta \alpha} \equiv \frac{1}{2} \left( \widetilde{S}^{\beta \alpha}_R - \widetilde{S}^{\beta \alpha}_L \right) \, ,
\end{equation}
for the flavor violating terms, and
\begin{equation}
  A^e_{\beta \beta} \equiv \left( \widetilde{S}_R^{\beta \beta} - \widetilde{S}_L^{\beta \beta} \right) \, ,
\end{equation}
for the flavor conserving ones.
With this translation, it is easy to check that both results agree up
to a global factor of $1/2$.

In order to compare the $\ell_\alpha^- \to \ell_\beta^- \ell_\beta^+ \ell_\beta^-$ decay widths obtained with both
parametrizations we need an expanded version of Eq.~\eqref{eq:widthS}
that includes terms up to $\mathcal{O}\left( \frac{m_\beta}{m_\alpha}
\right)$. This is given by
\begin{equation}
	\begin{split}
		& \Gamma_\phi\left( \ell_\alpha^- \to \ell_\beta^- \ell_\beta^+ \ell_\beta^- \right) =  \\
		& \frac{m_{\alpha}}{512 \pi^{3}} \Biggl\{ \left( \left| S_{L}^{\beta \alpha} \right|^{2} + \left| S_{R}^{\beta \alpha} \right|^{2} \right) \left\{ \left| S^{\beta \beta} \right|^{2} \left( 4 \log \frac{m_{\alpha}}{m_{\beta}} - \frac{49}{6} \right) - \frac{2}{6} \left[ \left( S^{\beta \beta *} \right)^2 + \left( S^{\beta \beta} \right)^2 \right] \right\} \Biggr. \\
		& - \frac{m_\alpha^2}{6} \biggl\{ S^{\beta \alpha}_L S^{\beta \beta} A^{S *}_{LL} +2 S_{L}^{\beta \alpha} S^{\beta \beta *} A_{L R}^{S *}+2 S_{R}^{\beta \alpha} S^{\beta \beta} A_{R L}^{S *}+S_{R}^{\beta \alpha} S^{\beta \beta *} A_{R R}^{S *}\biggr. \\		
		& - 12 \left( S^{\beta \alpha}_L S^{\beta \beta} A^{T *}_{LL} + S^{\beta \alpha}_R S^{\beta \beta *} A^{T *}_ {RR} \right) - 4 \left( S^{\beta \alpha}_R S^{\beta \beta} A^{V *}_{RL} + S^{\beta \alpha}_L S^{\beta \beta *} A^{V *}_{LR} \right) \\
		& + 6 e^2 \left( S^{\beta \alpha}_R S^{\beta \beta} K_2^{L *} + S^{\beta \alpha}_L S^{\beta \beta *} K_2^{R *} \right) - \frac{36 m_\beta}{m_\alpha} \left( S^{\beta \alpha}_R S^{\beta \beta} A^{T *}_{LL} + S^{\beta \alpha}_L S^{\beta \beta *} A^{T *}_{RR} \right) \\
		& + \frac{3 m_\beta}{2 m_\alpha} \biggl[ S^{\beta \beta} \left( 11 S^{\beta \alpha}_L A^{S *}_{RL} + 2 S^{\beta \alpha}_R A^{S *}_{LL} - 7 S^{\beta \alpha}_R A^{S *}_{LR} \right) \biggr. \\
		& + \biggl. S^{\beta \beta *} \left( 11 S^{\beta \alpha}_R A^{S *}_{LR} + 2 S^{\beta \alpha}_L A^{S *}_{RR} - 7 S^{\beta \alpha}_L A^{S *}_{RL} \right) \biggr] \\
		& - \frac{6 m_\beta}{m_\alpha} \left( S^{\beta \alpha}_L A^{S *}_{RL} - S^{\beta \alpha}_R A^{S *}_{LR} \right) \left( S^{\beta \beta} - S^{\beta \beta *} \right) \log \frac{m_{\alpha}}{m_{\beta}} \\
		& + \frac{12 m_\beta}{m_\alpha} \left( S^{\beta \alpha}_L A^{V *}_{RL} - 2 S^{\beta \alpha}_L A^{V *}_{RR} + 2 S^{\beta \alpha}_R A^{V *}_{LL} - S^{\beta \alpha}_R A^{V *}_{LR} \right) \left( S^{\beta \beta} - S^{\beta \beta *} \right) \log \frac{m_{\alpha}}{m_{\beta}} \\
		& + \frac{3 m_\beta}{m_\alpha} \biggl[ S^{\beta \beta} \left( -11 S^{\beta \alpha}_L A^{V *}_{RL} + 14 S^{\beta \alpha}_L A^{V *}_{RR} - 14 S^{\beta \alpha}_R A^{V *}_{LL} + 7 S^{\beta \alpha}_R A^{V *}_{LR} \right) \biggr. \\
		& + \biggl. S^{\beta \beta *} \left( -11 S^{\beta \alpha}_R A^{V *}_{LR} + 14 S^{\beta \alpha}_R A^{V *}_{LL} - 14 S^{\beta \alpha}_L A^{V *}_{RR} + 7 S^{\beta \alpha}_L A^{V *}_{RL} \right) \biggr] \\
		& - 4 \, e^2 \, \frac{m_\beta}{m_\alpha} \biggl\{ \left[ S^{\beta \alpha}_L \left(K_2^L\right)^{\beta \alpha \ast} + S^{\beta \alpha}_R \left(K_2^R\right)^{\beta \alpha \ast} \right] \left( S^{\beta \beta} + S^{\beta \beta *} \right) \left( 6 \log \frac{m_{\alpha}}{m_{\beta}} - \frac{21}{2} \right) \biggr. \\
		& \Biggl. + \, \biggl. \biggl. \left[ S^{\beta \alpha}_L S^{\beta \beta} \left(K_2^L\right)^{\beta \alpha \ast} + S^{\beta \alpha}_R S^{\beta \beta *} \left(K_2^R\right)^{\beta \alpha \ast} \right] \left( \pi^2 + 6 \log^2 2 - 6 \log^2 \frac{m_{\alpha}}{m_{\beta}} \right) \biggr\} + \cc \biggr\} \Biggr\} \, ,
	\end{split}
	\label{eq:widthSexp}
\end{equation}
where in this expression $A_{X Y}^I = \left( A_{X Y}^I \right)^{\beta
  \beta \beta \alpha}$. Replacing Eqs.~\eqref{eq:dic1} and
\eqref{eq:dic2} into Eq.~\eqref{eq:widthSexp} one finds full agreement
with Eq.~\eqref{eq:widthS2} to order $\mathcal{O}\left(
\frac{m_\beta}{m_\alpha} \right)$. This proves explicitly the
equivalence between both parametrizations in the calculation of
$\ell_\alpha^- \to \ell_\beta^- \ell_\beta^+ \ell_\beta^-$ mediated by
a pure pseudoscalar.

\bibliographystyle{utphys}
\bibliography{refs}

\end{document}